\documentclass[aps,prx,twocolumn,floatfix,superscriptaddress,longbibliography]{revtex4-2}
\usepackage{amsfonts, amsmath, amssymb}
\usepackage[compat=1.0.0]{tikz-feynman}
\usepackage{graphicx}
\usepackage{dcolumn}
\usepackage{bm}
\usepackage[normalem]{ulem}
\usepackage{color}
\usepackage[utf8]{inputenc}
\usepackage[colorlinks,citecolor=blue,linkcolor=blue,urlcolor=blue]{hyperref}
\usepackage{braket}
\usepackage[caption=false]{subfig}
\usepackage{color}
\usepackage{soul}
\usepackage{mathtools}
\usepackage{float}
\newcommand{\be}{\begin{equation}}
\newcommand{\ee}{\end{equation}}
\newcommand{\ba}{\begin{eqnarray}}
\newcommand{\ea}{\end{eqnarray}}

\newcommand{\sgn}{\mathop{\mathrm{sgn}}}

\include{physics}

  {\left\lbrace\begin{array}{@{}l@{}}}%
  {\end{array}\right.}

\begin{document}
\title{Cavity-mediated electron hopping in disordered quantum Hall systems}

\author{Cristiano Ciuti}
\affiliation{Universit\'{e} de Paris, Laboratoire Mat\'{e}riaux et Ph\'{e}nom\`{e}nes Quantiques (MPQ), CNRS-UMR7162, 75013 Paris, France}

\begin{abstract}
 We investigate the emergence of long-range electron hopping mediated by cavity vacuum fields in disordered quantum Hall systems. We show that  the counter-rotating (anti-resonant) light-matter interaction produces an effective hopping between disordered eigenstates within the last occupied Landau band. The process involves a number of intermediate states equal to the Landau degeneracy: each of these states consists of a virtual cavity photon and an electron excited in the next Landau band with the same spin. We study such a cavity-mediated hopping mechanism in the dual presence of a random disordered potential and a wall potential near the edges, accounting for both paramagnetic coupling and diamagnetic renormalization. We determine the cavity-mediated scattering rates, showing the impact on both bulk and edge states. The effect for edge states is shown to increase when their energy approaches the disordered bulk band, while for higher energy the edge states become asymptotically free.  We determine the scaling properties while increasing the Landau band degeneracy. Consequences on the quantum Hall physics and future perspectives are discussed.
\end{abstract}

\date{\today}
\maketitle

\section{Introduction}

The quantum Hall physics of a 2D electron gas immersed in a perpendicular magnetic field is one of the most fascinating chapters in the history of modern condensed matter physics \cite{Girvin_2019}. By tuning the magnetic field or the density of electrons, both of which control the filling factor of the single-particle Landau levels, a remarkable variety of electronic quantum phases can be obtained. In particular, a topologically robust quantization of the Hall charge trasverse conductance provides the most stable resistance standard known in metrology \cite{VK_1980,Jeanneret_2001}.

Another prominent branch of quantum physics
is Quantum Electrodynamics (QED). In low energy physics, QED is at the heart of celebrated vacuum effects such as the atomic Lamb spectral shift as well as the Casimir and van der Waals  forces \cite{milonni1994the}. 
 Cavity QED\cite{Haroche} can enhance vacuum effects by increasing light-matter interaction via tight spatial field confinement of quantum  modes in properly engineered electromagnetic resonators. Originally born in the context of atomic physics, cavity QED has become an exciting research field in solid-state systems \cite{Hennessy2007} and superconducting quantum circuits (circuit QED) \cite{Blais2021}.  
 
 In recent years, there has been a growing interest in the regime of ultrastrong light-matter interaction \cite{Forn_2019,Frisk_Kockum_2019}, which is achieved when the coupling between a photon and an elementary electronic transition becomes comparable to the photon and transition frequencies \cite{Ciuti_2005}. In particular, such a regime was predicted for the coupling of the cyclotron transition of a 2D electron gas to a cavity mode \cite{Hagenmuller_2010} and experimentally demonstrated by using deeply subwavelength THz split-ring resonators \cite{Scalari_2012}.  Interesting  linear and nonlinear optical properties of the related Landau polaritons have been investigated in a recent series of experimental spectroscopy works \cite{Maissen2014,Zhang2016,Bayer2017,Li2018,Halbhuber2020,Mornhinweg_2021}. Other investigations have instead exploited optics as a probe of electronic quantum Hall physical properties \cite{Smolka2014,Hafezi2017,Ravets2018,Knppel2019},  or optical pumping as a way to manipulate electronic quantum Hall states \cite{Hafezi_2021}. We also wish to highlight a third optical research direction aimed at the realization of quantum Hall states of light in purely photonic systems \cite{Ozawa_2019,De_Bernardis_2021}. 
  
  In the broader context, an emerging field is currently focused on the manipulation of matter by vacuum fields in physics and chemistry \cite{Garcia_2021}. The modification of electron transport by a passive cavity (no illumination) has been studied for organic disordered materials \cite{Orgiu2015,Nagarajan2020,Hagenmuller2017,Hagenmuller2018,Botzung_2020,Chavez2021} and for 2D electron gases in the semiclassical Shubnikov-de Haas magnetotransport regime \cite{ParaviciniBagliani2018,BC_2018}, as well as for the vertical transport in semiconductor heterostructures \cite{NaudetBaulieu2019}.  A recent study has also proposed cavity-mediated superconductivity of a 2D electron gas \cite{Schlawin_2019}, where the electron pairing mechanism is based on the exchange of virtual cavity photons. Remarkably, recent pioneering experiments \cite{Appugliese_2021} on high-mobility 2D electron gases have shown that the quantum Hall transport can be dramatically affected by a cavity resonator without illumination with a breakdown of the topological protection and a non-trivial modification of both transverse and longitudinal resistance in the integer quantum Hall regime (the fractional quantum Hall features have been shown to be instead largely immune to the cavity). The physics of cavity-controlled quantum Hall systems is in its very infancy and provides an intriguing platform for exploring the manipulation of electronic properties by vacuum fields.
                
 In this article, we present a microscopic theory revealing how the electromagnetic vacuum fields of a cavity mode can mediate long-range electron hopping processes between disordered eigenstates in a quantum Hall system.  We show that these processes are due to the counter-rotating (anti-resonant) light-matter interaction via the exchange of a virtual cavity photon.  This paper is organized as follows. In Sec. \ref{framework}, we present the general theoretical framework, describing Landau electronic states in the presence of a disordered electronic potential, an edge wall potential and a spatially homogeneous quantum electromagnetic cavity mode. We consider the microscopic Hamiltonian including the paramagnetic and diamagnetic contributions, expressing it in a compact form in terms of the disordered eigenstates and the photon mode renormalized by the diamagnetic interaction. 
 In Sec. \ref{hopping}, we derive the cavity-mediated electron hopping in terms of the disordered eigenstates and the corresponding cavity-mediated scattering rates.
 In Sec. \ref{finite-size}, we report finite-size numerical calculations and find the scaling properties in the limit of large number of electrons.  In Sec. \ref{Discussion}, we discuss consequences on quantum Hall physics. In Sec. \ref{conclusions}, we draw the conclusions and perspectives of this work.

\section{Theoretical framework}
\label{framework}
\begin{figure}[t!]
	\centering
	\includegraphics[scale=0.17]{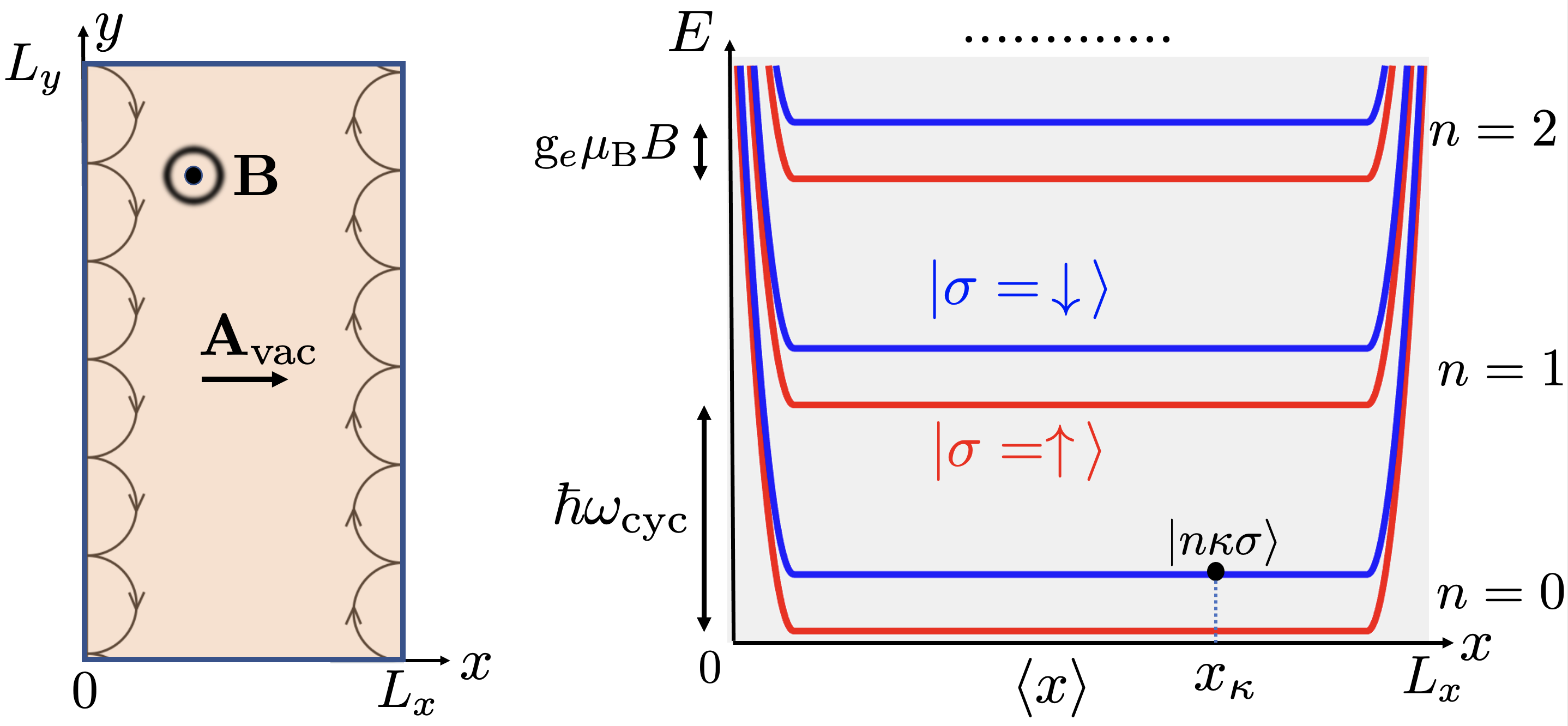} 
	\caption{Left: scheme of the considered 2D electron system in a rectangular geometry with a perpendicular static magnetic field $\mathbf B$. The electronic system is coupled to a quantum electromagnetic mode with photon energy $\hbar \omega_{\rm cav}$ and with a vacuum vector potential $\mathbf{A}_{\rm vac}$ linearly polarized along the $x$ direction. Counter-propagating Landau edge states are pictorially represented by skipping orbits. Right: a sketch of the bare (no disorder) energy of the Landau single-particle eigenstates $\vert n \kappa \sigma \rangle$ including the Zeeman spin splitting and a smooth wall potential at the edges along the $x$ direction. In the chosen Landau gauge, the orbit center position $x_{\kappa}$ is proportional to the orbital quantum number $\kappa \in \{ 1,2,...,N_{\rm deg} \}$, where $N_{\rm deg}$ is the Landau level orbital degeneracy. }
	\label{sketch}
\end{figure}

Let us consider a 2D electron gas that is subject to a perpendicular magnetic field $B$ in a rectangular geometry, as depicted in Fig. \ref{sketch}.  We will consider a wall potential $W(x)$ near the edges \cite{Halperin_1981}.
In the second quantization formalism, the bare electronic energy of the Landau levels including the wall potential and the Zeeman contribution is described by the Hamiltonian: 
\begin{equation}
\hat{H}_{\rm el}  = \sum_{n, \kappa, \sigma}   (E_{n,\sigma} + W_{n,\kappa}) \hat{c}^{\dagger}_{n,\kappa,\sigma} \hat{c}_{n,\kappa,\sigma}\,,
\end{equation}
where the Landau  energies are $ E_{n, \sigma} = 
E_n -  \frac{1}{2} \sigma \, {\rm g}_e \mu_{\rm B} B $ with $E_n = n \hbar \omega_{\rm cyc}$. The cyclotron frequency  is given by $\omega_{\rm cyc} = e B/m$ ($m$ is the effective electron mass) and the electron Zeeman splitting is ${\rm g}_{\rm e} \mu_{\rm B} B$, where ${\rm g}_{\rm e}$ is the effective gyromagnetic factor and $\mu_{\rm B}$ the Bohr magneton. The operator
  $\hat{c}^{\dagger}_{n,\kappa,\sigma}$ creates an electron in the state with orbital quantum numbers $n \in \{0, 1, 2,...\}$, $\kappa \in \{1,2,...,N_{\rm deg} \}$ and with $\sigma \in \{\uparrow ,\downarrow \}$ the spin projection along the $z$-direction. Each Landau band has an orbital degeneracy equal to $N_{\rm deg} = L_x L_y/(2 \pi l_{\rm cyc}^2)$, where the cyclotron length is $l_{\rm cyc} = \sqrt{\frac{\hbar}{m \omega_{\rm cyc}}}$. In the chosen Landau gauge, the classical vector potential generating the static magnetic field bias is ${\mathbf A} = B x \, {\mathbf e}_y$, where ${\mathbf e}_y$ is the unit vector pointing in the $y$-direction. 
  The Landau states in the presence of the wall potential have wavefunctions $\Psi_{n,\kappa} ({\mathbf r}) = \langle {\bf r}\vert n \kappa \rangle =  {\mathcal N}_n F_{n}(\frac{x-\tilde{x}_{\kappa}}{l_{\rm cyc}}) e^{i \frac{2 \pi \kappa y}{L_y}}$, where  the normalization factor is ${\mathcal N}_n = \frac{1}{\sqrt{\sqrt{\pi} \,2^n n! \, l_{\rm cyc} L_y }}$ and the function $F_n(\xi) = H_n(\xi) e^{-\xi^2/2}$ depends on the Hermite polynomial $H_n$ of order $n$. The Landau orbit center positions are given by the expression $\tilde{x}_{\kappa}  =x_{\kappa} + \delta x_{\kappa}$ where $x_{\kappa}  =  2 \pi  \frac{l_{\rm cyc}^2}{L_y} \kappa$. For a smooth wall potential, we have $\delta x_{\kappa} \simeq -W'(x_{\kappa})/(m \omega_{\rm cyc}^2)$ and $W_{n,\kappa} \simeq W(x_{\kappa})$.
  The Landau states in presence of the wall potential acquire a finite velocity along the $y$ direction, namely $v^{(y)}_{\kappa} = W'(x_{\kappa})/(m \omega_{cyc})$.

In the following, we will assume that the system has some moderate static disorder coupling Landau states with the same orbital quantum number $n$ (we will neglect Landau level mixing):
\begin{equation}
 \hat{H}_{\rm dis} = \sum_{n, \kappa,\kappa'} V^{(n)}_{\kappa,\kappa'}
 \hat{c}^{\dagger}_{n,\kappa,\sigma} \hat{c}_{n,\kappa',\sigma}\,.
 \end{equation}

 Now, let the 2D electron gas be coupled to the quantum mode field of an electromagnetic resonator with frequency $\omega_{\rm cav}$ and a spatially homogeneous mode polarized along the $x$ direction, represented by the vector potential operator $\hat{\mathbf{A}}_{\rm vac} =  {A}_{\rm vac} \,{\bf{e}_x} \, (\hat{a}+\hat{a}^{\dagger})$.
 This is a configuration close to what achieved in the capacitive spatial gap of split-ring resonators \cite{ParaviciniBagliani2018}.
 The bare cavity Hamiltonian is  $\hat{H}_{\rm cav} = \hbar \omega_{\rm cav} \hat{a}^{\dagger} \hat{a} \, $,  where $\hat{a}^{\dagger}$ is the photon creation operator. Using the Coulomb gauge and the minimal coupling Hamiltonian for the electrons, the light-matter interaction  has the following paramagnetic contribution \cite{BC_2018}:
\begin{equation}
\hat{H}_{\rm para} = \sum_{n,\kappa,\sigma} (- {\rm i})
\hbar g \sqrt{n+1} \, (\hat{a}+\hat{a}^{\dagger})
 \hat{c}^{\dagger}_{n+1,\kappa,\sigma} \hat{c}_{n,\kappa,\sigma}  + 
\mathrm{h.c.}
 \,, \label{para_clean}
\end{equation}
where the vacuum Rabi frequency is defined by the relation
\begin{equation}
g = \frac{e A_{\rm vac}}{\hbar}	\sqrt{\frac{\hbar \omega_{cyc}}{2 m}}\, .
\end{equation}
In addition, there is a diamagnetic contribution
\begin{equation}
\hat{H} _{\rm dia}= 
N_{\rm el} \frac{e^2 A_{\rm vac}^2}{2 m} 
(\hat{a}+\hat{a}^{\dagger})^2  \, = \frac{\hbar\Omega^2}{\omega_{\rm cyc}} (\hat{a}+\hat{a}^{\dagger})^2 \,,
\end{equation}
where we have introducted the collective Rabi frequency 
\begin{equation}
\Omega  = g \sqrt{N_{\rm el}}
\end{equation}
 with $N_{\rm el}$ the total number of electrons. The additional Hamiltonian term is the Coulomb interaction that will not be considered in this work.

\begin{figure*}[t!]
	\centering
	\includegraphics[scale=0.28]{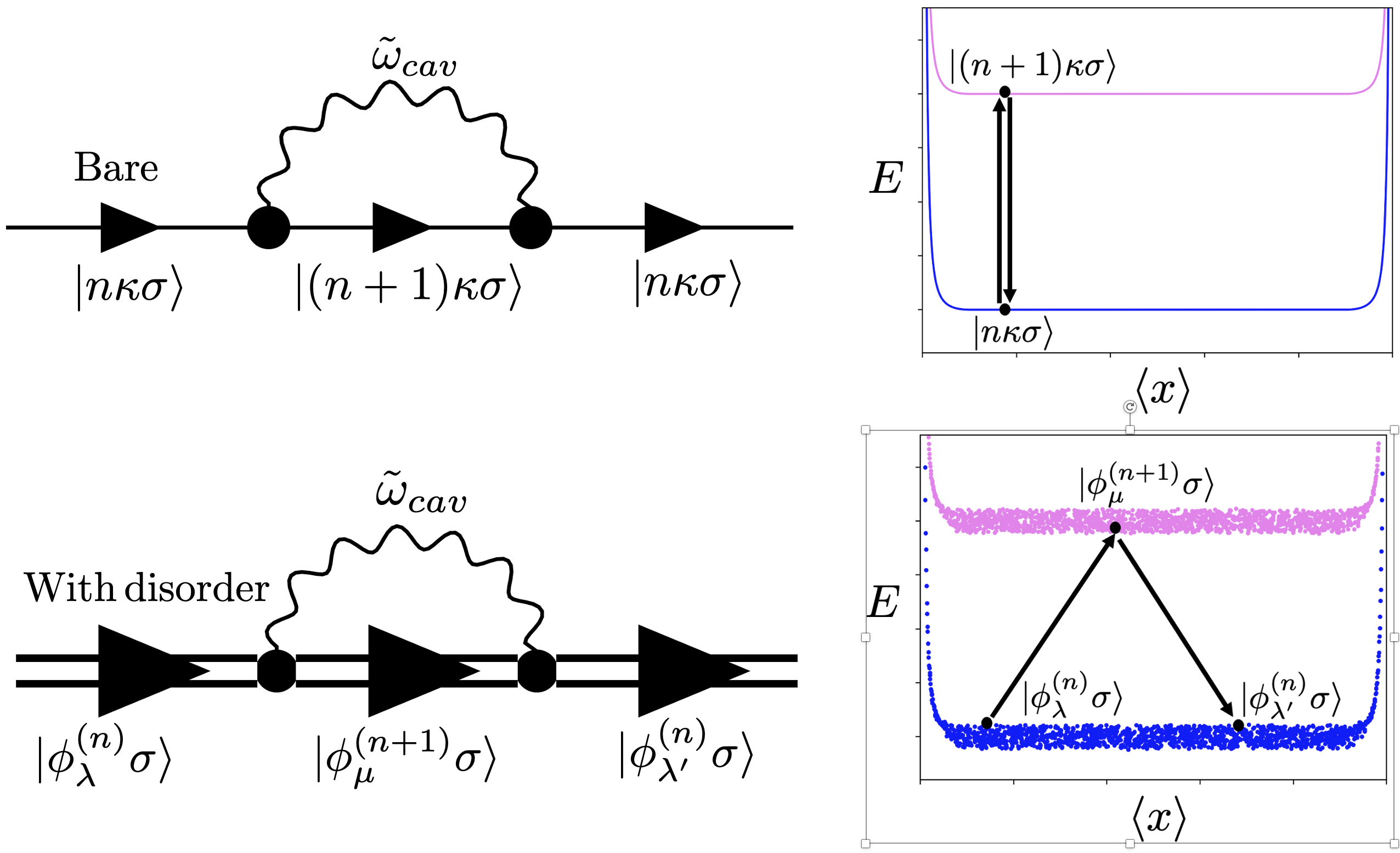} 
	%
	\caption{Top: on the left, a diagram represents an electron initially in a bare (no disorder) Landau state $\vert n \kappa \sigma \rangle$ of the last occupied Landau band. The intermediate state consists of a virtual cavity photon and the electron promoted to the $(n+1)$-band.  This process is due to the anti-resonant light-matter interaction. Due to the selection rules of the cavity coupling, both the Landau quantum number $\kappa$ (hence the orbit center position $x_{\kappa}$) and the spin are conserved. The final state of the process can be only the initial one. On the right, the same process depicted on the plot of the energy versus the average $x$ position.
	Bottom: analogous propagation process for an electron initially in a disordered single-particle eigenstate $\vert \phi^{(n)} _{\lambda} \sigma \rangle$ where $\lambda$ is the quantum number labeling the disordered eigenstate.
	 Via the intermediate state involving a virtual cavity photon, the initial state is coupled to a different final state $\vert \phi^{(n)} _{\lambda'} \sigma \rangle$. Indeed, in presence of disorder the quantum number $\lambda$ is not conserved by the light-matter interaction. This cavity-mediated process is an effective hopping between the disordered state $\vert \phi^{(n)} _{\lambda} \sigma \rangle$ and the state $\vert \phi^{(n)} _{\lambda'} \sigma \rangle$.  Note that the number of intermediate states $\vert \phi^{(n+1)} _{\mu} \sigma \rangle$ is equal to the Landau orbital degeneracy $N_{\rm deg}$. On the right, the same process depicted on the plot of the electronic energy versus the average position.} 
	\label{diagrams}
\end{figure*}
The cavity Hamiltonian $\hat{H}_{\rm cav}$ and the diamagnetic term $\hat{H}_{\rm dia}$ depend only on the photon operators and are quadratic with respect to them.
Hence, via a Bogoliubov transformation, we get the diagonal form:
\begin{equation}
\hat{H}_{\rm mode} =\hat{H}_{\rm cav} + \hat{H}_{\rm dia} = \hbar \tilde{\omega}_{\rm cav}  \hat{\alpha}^{\dagger} \hat{\alpha} + {\rm const.}
\label{mode}
\, , \end{equation}
describing a boson mode with renormalized frequency
\begin{equation}
\tilde{\omega}_{cav}  = \sqrt{\omega_{\rm cav}^2 +  4 \frac{ \Omega^2}{\omega_{\rm cyc}} \omega_{\rm cav}} \, .
\end{equation} 
The dressed bosonic photon operator reads
\begin{equation}
\hat{\alpha}
= \frac{1}{2 \sqrt{\tilde{\omega}_{\rm cav} \omega_{\rm cav}}}
\left [\left ( \tilde{\omega}_{\rm cav} + \omega_{\rm cav}  \right )  \hat{a} + 
\left ( \tilde{\omega}_{\rm cav} -  \omega_{\rm cav}  \right )  \hat{a}^{\dagger} \right ] \, .
\end{equation}

The bare Landau single-particle Hamiltonian with the disorder potential and the wall potential can be diagonalized in the form:
\begin{equation}
\hat{H}_{\rm sp} =\hat{H}_{\rm el} + \hat{H}_{\rm dis}  =
\sum_{n, \kappa, \sigma} \left (  \epsilon_{n,\lambda} + 
\frac{1}{2} \sigma \, {\rm g}_e \mu_{\rm B} B \right ) \hat{d}^{\dagger}_{n,\lambda,\sigma} \hat{d}_{n,\lambda,\sigma} \, ,
\label{sp}
\end{equation}
where $\epsilon_{n,\lambda}$ are the energies of the single-particle orbital eigenstates of the  Landau band with orbital quantum number $n$, namely:
\begin{equation}
\vert \phi^{(n)}_{\lambda} \rangle = \sum_{\kappa} \langle n \kappa \vert \phi^{(n)}_{\lambda} \rangle  \vert n \kappa \rangle \, .
\end{equation}
The corresponding  fermionic  operators are given by the relation $
\hat{c}_{n,\kappa ,\sigma} = \sum_{\lambda}   \langle n \kappa \vert  \phi^{(n)}_{\lambda} \rangle  \,  \hat{d}_{n,\lambda,\sigma} $.

Since  $
\hat{a} + \hat{a}^{\dagger}
=   \sqrt{\omega_{cav}/\tilde{\omega}_{cav} }
( \hat{\alpha} + \hat{\alpha} ^{\dagger})$,
 the paramagnetic coupling can be rewritten in terms of renormalized photon mode operators and disordered eigenstates as:
\begin{equation}
\hat{H}_{\rm para} = \sum_{n,\lambda,\mu,\sigma} (-{\rm i})
\hbar \tilde{g}^{(n,n+1)}_{\lambda,\mu}  \, (\hat{\alpha}+\hat{\alpha}^{\dagger})
 \,  \hat{d}^{\dagger}_{n+1,\mu,\sigma} \hat{d}_{n,\lambda,\sigma}  + 
{\mathrm{h.c.}}
\,, \label{coupl}
\end{equation}
where the coupling constant is:
\begin{equation}
\tilde{g}^{(n,n+1)}_{\lambda,\mu} = \tilde{g}  \sqrt{n+1} \sum_{\kappa} 
\,  \langle  \phi^{(n+1)}_{\mu} \vert  n +1  \, \kappa  \rangle  \langle n\kappa \vert  \phi^{(n)}_{\lambda} \rangle 
\,, \label{dis_g}
\end{equation}
with
\begin{equation}
\tilde{g}
=   g \sqrt{\omega_{cav}/\tilde{\omega}_{cav} } \,.
\end{equation}.

In conclusion, we have recast the total Hamiltonian in a much simpler form, namely
\begin{equation}
\hat{H} = \hat{H}_{\rm sp} + \hat{H}_{\rm mode} + \hat{H}_{\rm coupl},
\end{equation}
where the single-particle Hamiltonian $\hat{H}_{\rm sp}$ is given by Eq. (\ref{sp}), the diamagnetically-renormalized mode Hamiltonian  $\hat{H}_{\rm mode}$ is given by Eq. (\ref{mode}). Finally, the paramagnetic coupling $\hat{H}_{\rm para}$, expressed in terms of the renormalized boson mode and disordered eigenstates, is given by Eq. (\ref{coupl}).
Note that the single-electron vacuum Rabi frequency $\tilde{g}^{(n,n+1)}_{\lambda,\mu}$ depends on the disorder eigenstates and is renormalized by the diamagnetic interaction. 

\section{Cavity-mediated hopping}
\label{hopping}
As expressed in Eq. (\ref{para_clean}), in the absence of disorder the paramagnetic interaction conserves both the spin $\sigma$ and the orbital quantum number $\kappa$. With disorder, spin is still conserved, but, as shown in Eqs. (\ref{coupl}) and (\ref{dis_g}), the situation is radically different.  As depicted in Fig. \ref{diagrams},  the key point is that the counter-rotating (anti-resonant) terms of the paramagnetic interaction can couple a generic disorder eigenstate $\vert \phi^{(n)}_{\lambda} \rangle$  to any other disordered eigenstate  $\vert \phi^{(n)}_{\lambda'} \rangle$ via an intermediate virtual excited state. Indeed, an electron occupying the state $\vert \phi^{(n)}_{\lambda} \rangle$ can be promoted to the state $\mu$ in the $(n+1)$-band with the simultaneous creation of a photon with energy $\hbar \tilde{\omega}_{cav}$. The Hamiltonian matrix element for such a virtual process is $(-{\rm i}) \tilde{g}^{(n,n+1)}_{\lambda,\mu}$ and the corresponding energy penalty is
$\epsilon_{n,\lambda} - \epsilon_{n+1,\mu} - \hbar \tilde{\omega}_{cav}$. Via the reverse counter-rotating process, the photon can be re-absorbed and the electron demoted back to the $n$-band, but in a different and unoccupied disordered state $\lambda'$. The Hamiltonian matrix element for such process is 
${\rm i} \tilde{g}^{(n,n+1)\, \star}_{\lambda',\mu}$. 
Using perturbation theory, the effective coupling \cite{Malrieu_1985,Effective_2002} of the state $\lambda$ and to the state $\lambda'$ in the $n$-band can be approximated by the expression:
\begin{equation}
\Gamma^{(n)}_{\lambda, \lambda'} \simeq  \sum_{\mu} \frac{ \hbar^2 \, \tilde{g}^{(n,n+1)}_{\lambda,\mu} \tilde{g}^{(n,n+1)\, \star}_{\lambda',\mu}}
{\epsilon_{n, \lambda}- \epsilon_{n+1,\mu} - \hbar \tilde{\omega}_{\rm cav}}\, ,
\end{equation}
which has been obtained by summing over all possible intermediate states ${\mu}$. Importantly, the number of intermediate states is exactly equal to the macroscopic Landau degeneracy $N_{\rm deg}$.  This perturbative formula holds as long as 
$
\epsilon_{n+1,\mu} + \hbar \tilde{\omega}_{cav} \gg \epsilon_{n,\lambda} \,,
$ which is easily fulfilled when $\epsilon_{n,\lambda} < E_{n+1}$.

Note that for a given pair of states $\lambda$ and $\lambda'$ in the $n$-band, the dependence of $\Gamma^{(n)}_{\lambda, \lambda'}$ on the total number of electrons $N_{\rm el}$ and hence on the filling factor $\nu = N_{\rm el}/N_{\rm deg}$ enters only in the diamagnetic renormalization of the cavity mode frequency $\omega_{\rm cav}$ (replaced by $\tilde{\omega}_{\rm cav}$) and  of the bare vacuum Rabi frequency $g$ (replaced by $\tilde{g}$).

Having determined the cavity-mediated hopping coupling, we can evaluate the corresponding scattering rates with the Fermi golden rule:

\begin{equation}
\frac{1}{\tau^{(sc)}_{n,\lambda}} = \frac{2 \pi}{\hbar} \sum_{\lambda' \neq \lambda} \vert \Gamma^{(n)}_{\lambda, \lambda'} \vert^2 \delta (\epsilon_{n,\lambda} - \epsilon_{n,\lambda'} )
\, . \label{rates}
	\end{equation}

Of course, this formula is applicable in the case when the final states of the cavity-mediated scattering process are unoccupied.

If we are interested in transport properties when the current flows along the $y$ direction, a relevant quantity is the velocity $v^{(y)}_{n,\lambda}$ of the disordered eigenstates. 
This is given by the following expression: 

\begin{equation}
v^{(y)}_{n,\lambda} = \sgn(v^{(y)}_{n,\lambda})\frac{L_y}{\tau^{(tr)}_{n,\lambda}} = \sum_{\kappa} \vert \langle \phi^{(n)}_{\lambda} \vert n \kappa \rangle \vert^2 v^{(y)}_{\kappa}
\, , \label{speed}
\end{equation}
where we have also introduced $\tau^{(tr)}_{n,\lambda}$, which is the time for an electron to transit the channel length $L_y$ when populating the state $\vert \phi^{(n)}_{\lambda} \rangle$.  	
\begin{figure}[ht!]
	\centering
	\includegraphics[scale=0.45]{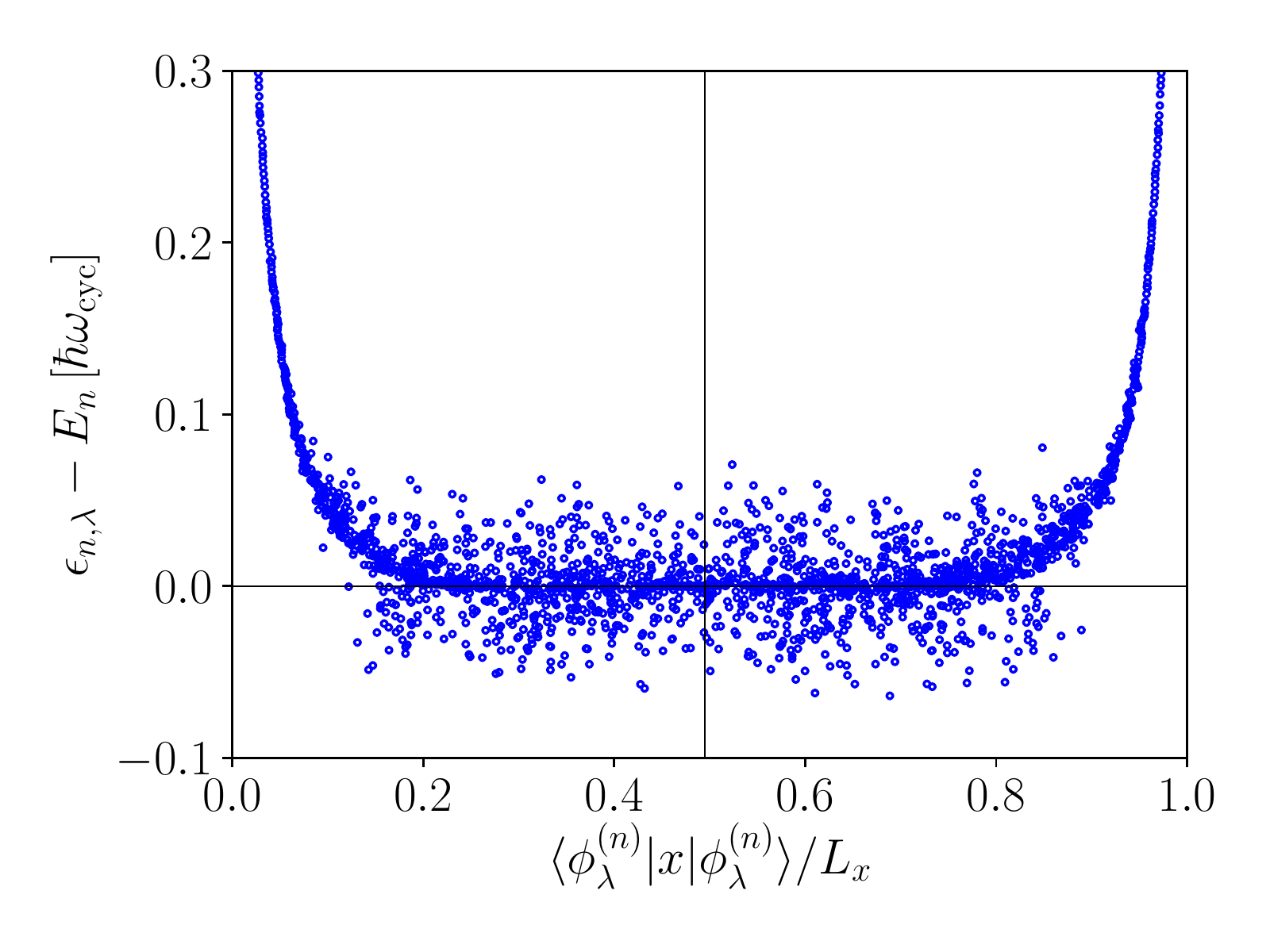}  \\
	\includegraphics[scale=0.45]{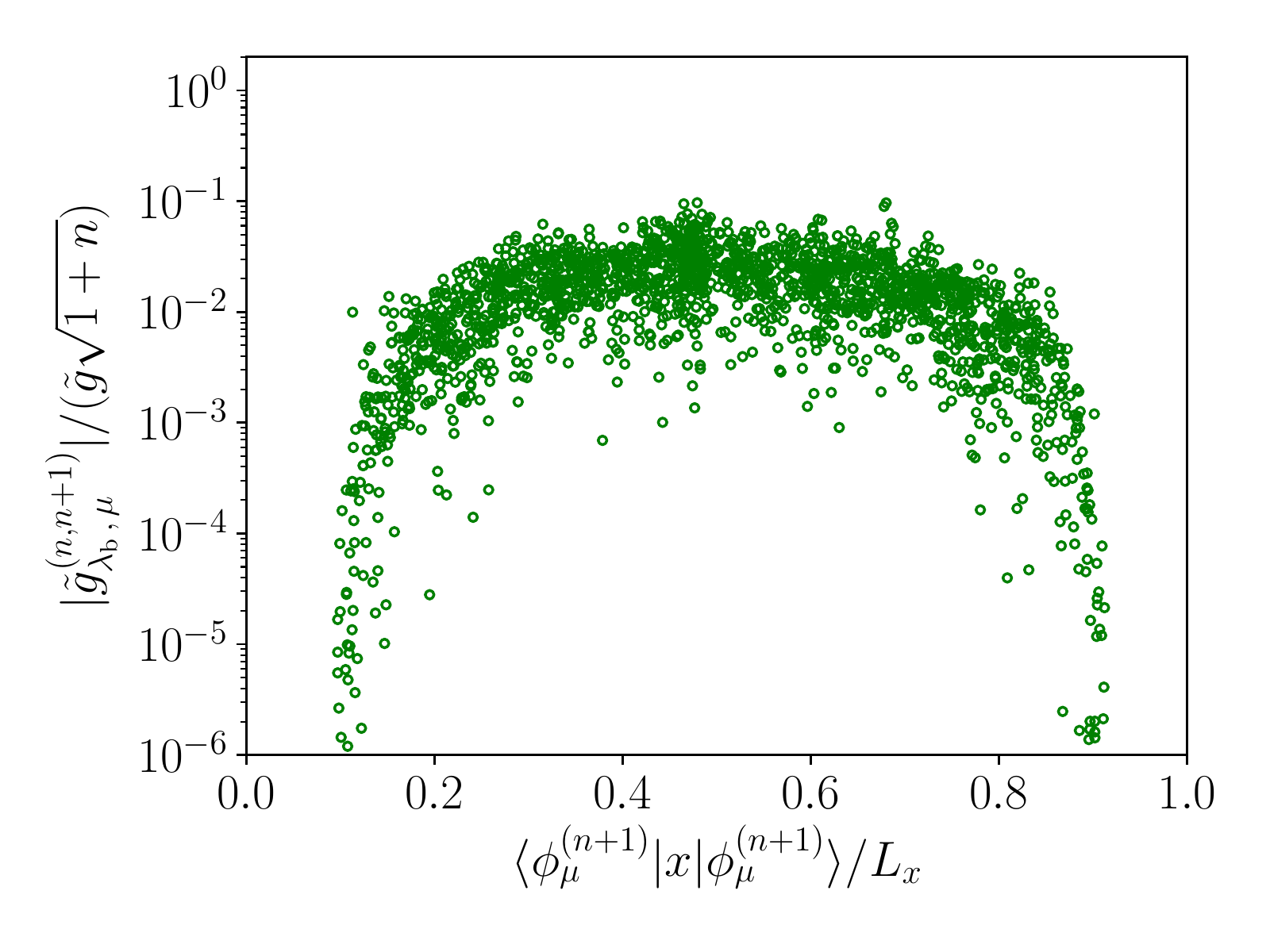} \\
	\includegraphics[scale=0.45]{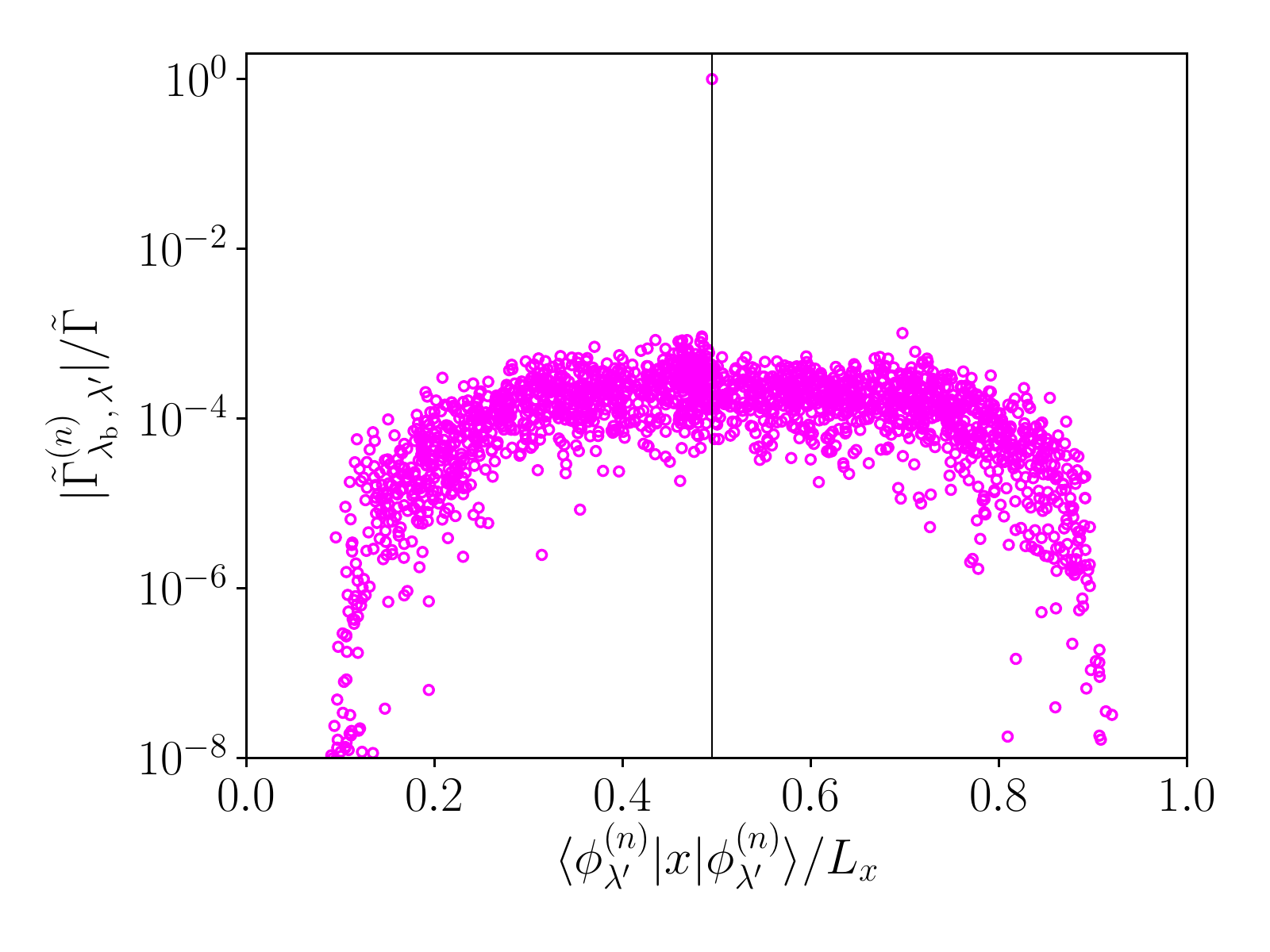} \\
	\caption{Top panel: energies $\epsilon_{n,\lambda}$ versus the average position $x$ of the corresponding disordered eigenstates. The horizontal and vertical lines indicate energy and average  position for a bulk state $\lambda_{\rm b}$. Middle panel: vacuum Rabi coupling frequency $\vert \tilde{g}^{(n,n+1)}_{\lambda_{\rm b},\mu} \vert$ between the considered state and the states
		$\vert \phi^{(n+1)}_{\mu} \rangle$ versus their average position.
		Bottom panel: cavity-mediated hopping coupling between $\vert \phi^{(n)}_{\lambda_{\rm b}} \rangle$ and the states $\vert \phi^{(n)}_{\lambda'} \rangle$ versus their average position. The coupling is normalized to  $\tilde{\Gamma}$ defined in Eq. (\ref{E_ref}).
		Parameters: $N_{\rm deg} = 2400$, $n=4$, $L_x = 10 \mu$m, $L_{\rm e} = 2.5 \mu$m, $r= 0.9$,  $V_{\rm e} = 0.03 \hbar \omega_{\rm cyc}$, ${\mathcal V}^{({\rm imp})}_{\rm max}  =  1.5 \cdot 10^{-5} \hbar \omega_{\rm cyc} L_x L_y$, $N_{\rm imp} = 2000$, $B = 0.795 \,$T, $m = 0.067 m_0$ ($m_0$ is the  electron mass). For the bottom panel, $N_{\rm el} = 24000$, $g = 0.0051 \,\omega_{\rm cyc}$ and $\omega_{\rm cav} = 0.39 \, \omega_{\rm cyc}$. For these parameters, $\tilde{g} \simeq 0.61 g$.	\label{example_bulk} }
\end{figure}

\begin{figure}[ht!]
	\centering 
	\includegraphics[scale=0.45]{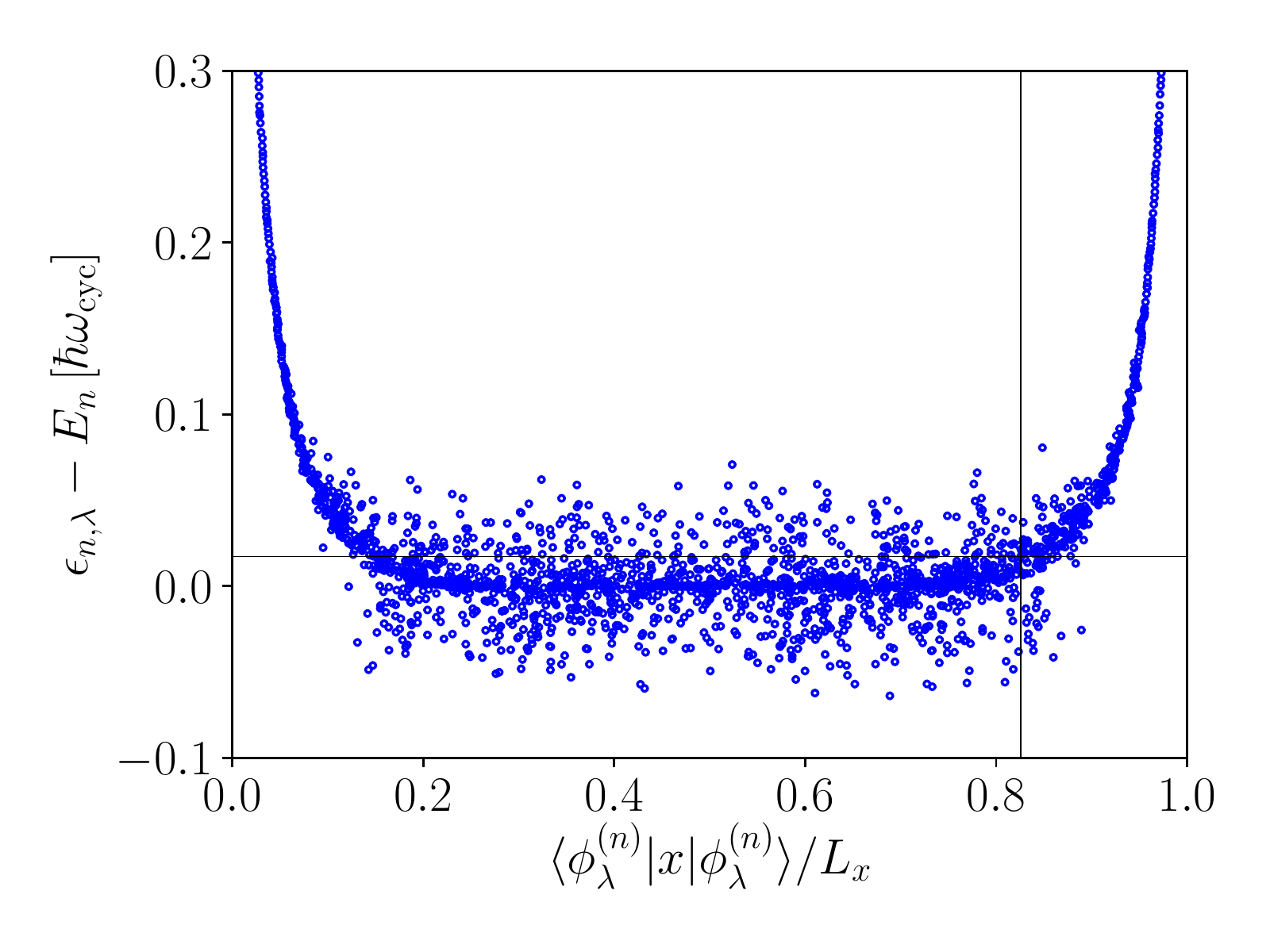}  \\
	\includegraphics[scale=0.45]{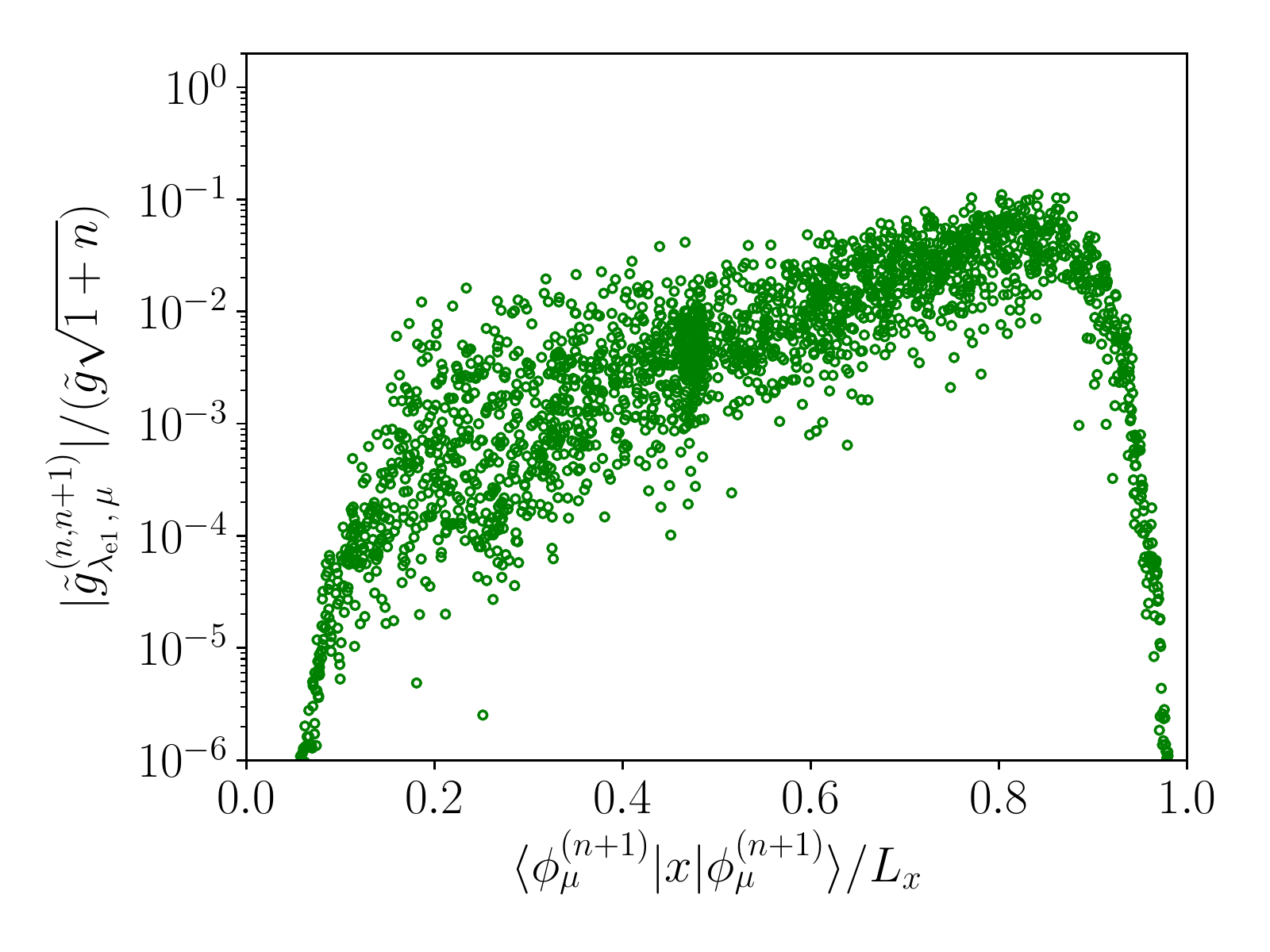} \\
	\includegraphics[scale=0.45]{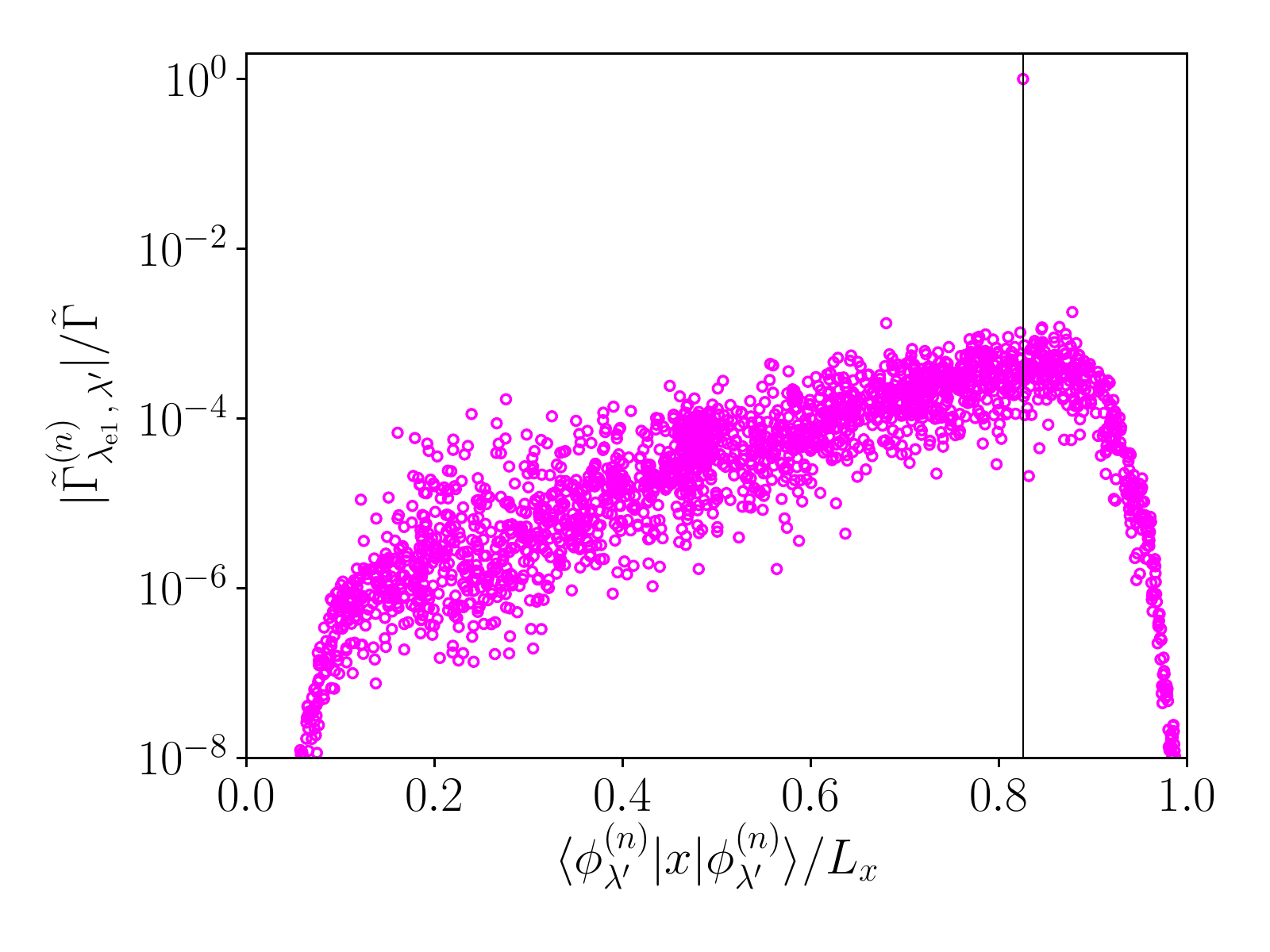} 
	\caption{Same as Fig. \ref{example_bulk}, but for an edge state $\lambda_{{\rm e}1}$ energetically close to the disordered bulk band. The energy and average position $x$ is indicated by the vertical and horizontal lines in the top panel. 	\label{example_edge_1}	}
\end{figure}

\begin{figure}[ht!]
	\centering 
	\includegraphics[scale=0.45]{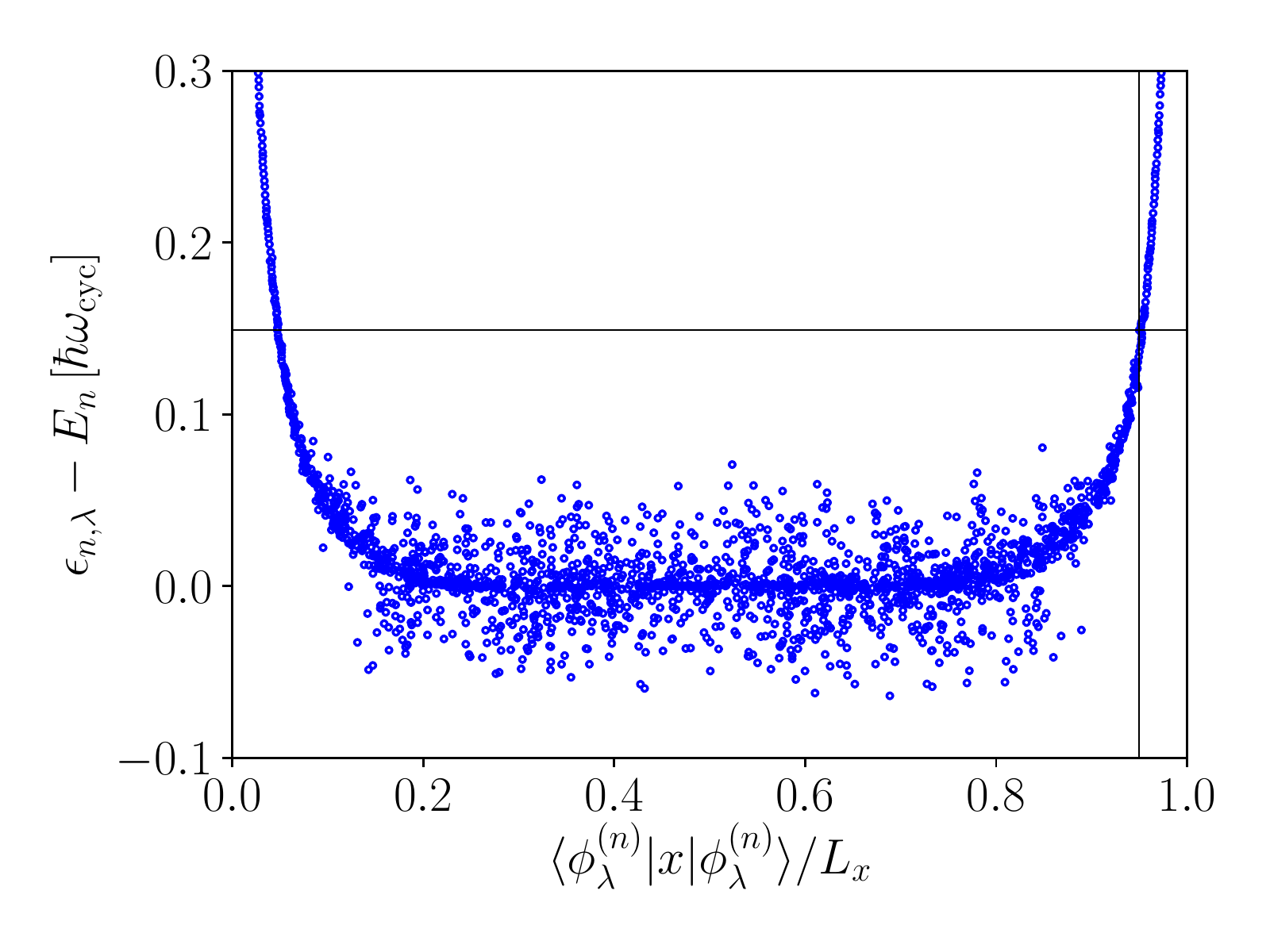}  \\
	\includegraphics[scale=0.45]{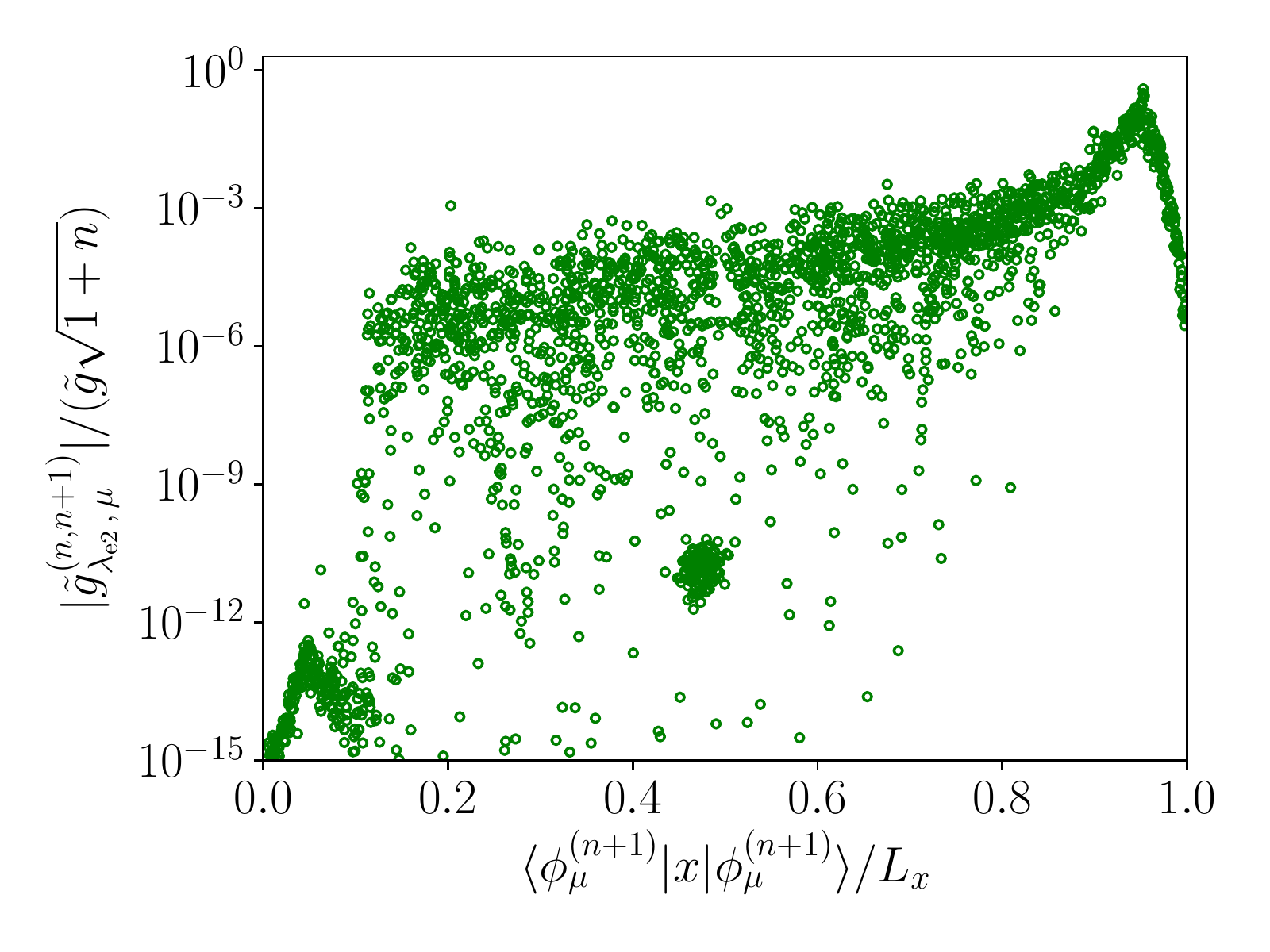} \\
	\includegraphics[scale=0.45]{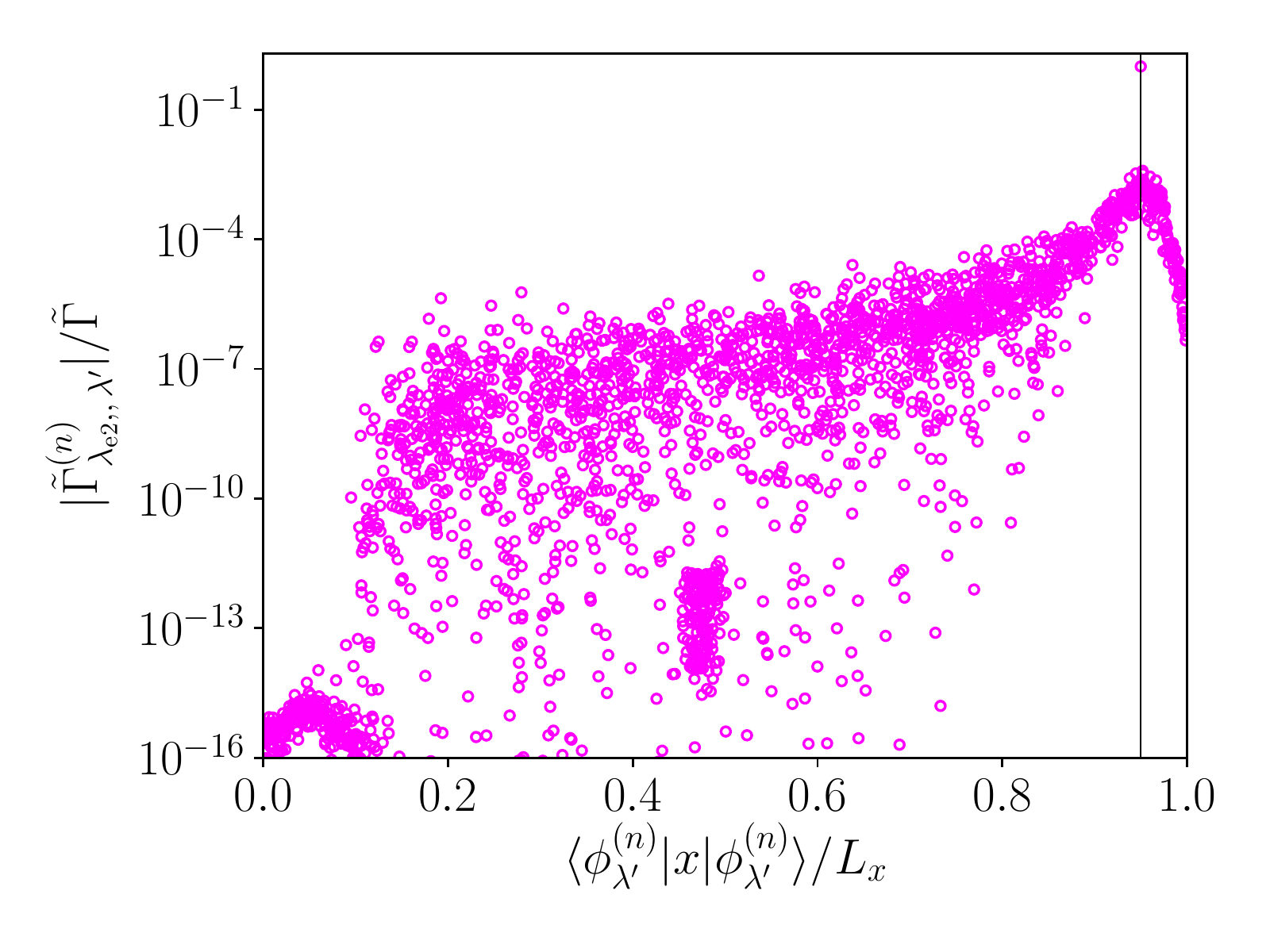} 
	\caption{Same as Fig. \ref{example_bulk}, but for another edge state $\lambda_{{\rm e}2}$, whose energy and average position $x$ is indicated by the vertical and horizontal lines in the top panel. This edge state has a much higher energy compared to that in Fig. \ref{example_edge_1} and the cavity-mediated hopping is strongly suppressed (note the very different log scale for the vertical axis in the middle and bottom panels). 	\label{example_edge_high}}
	
\end{figure}

\begin{figure*}[ht!]
	\centering 
	\includegraphics[scale=0.3]{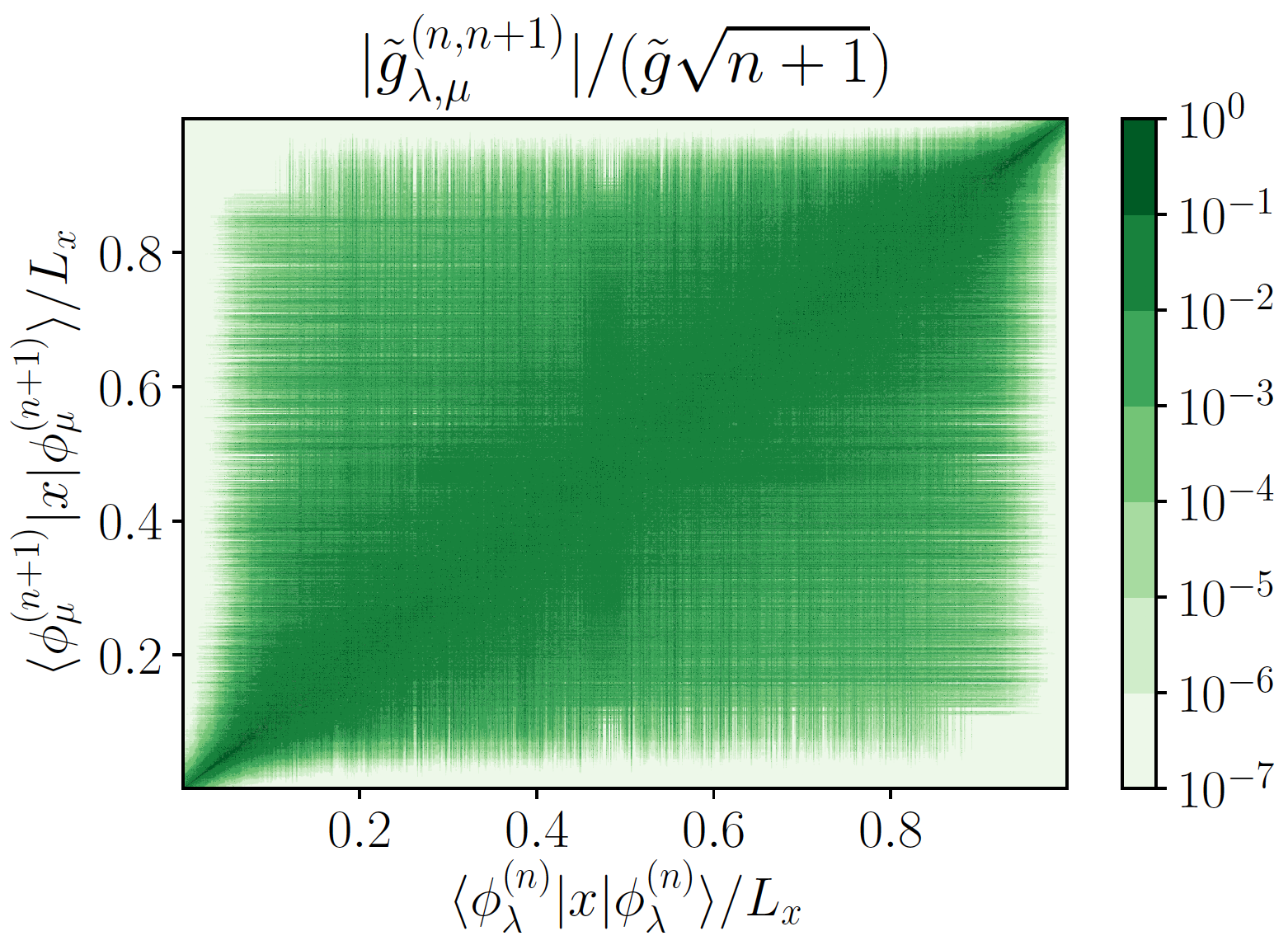}  
	\includegraphics[scale=0.3]{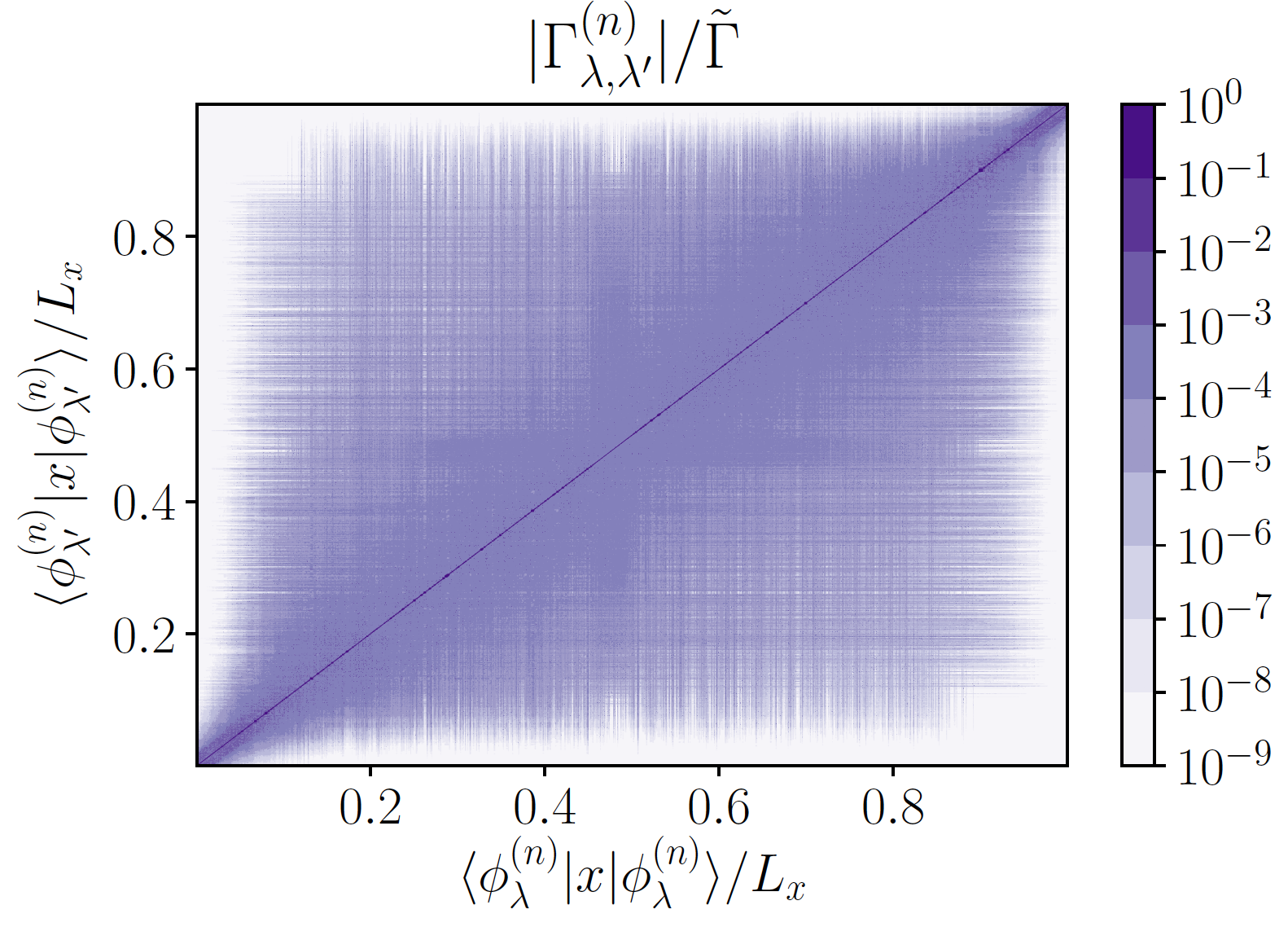} 
	\caption{Left panel: color plot of the normalized vacuum Rabi frequency (see title) as a function of the average position $x$ of the disordered eigenstates $\vert \phi^{(n)}_{\lambda} \rangle$ and $\vert \phi^{(n+1)}_{\mu} \rangle$. Right panel: color plot of the normalized cavity-mediated hopping (see title) as a function of the average position $x$ of the states $\vert \phi^{(n)}_{\lambda} \rangle$ and $\vert \phi^{(n)}_{\lambda'} \rangle$. Same parameters as in Fig. \ref{example_bulk}.	
		 	\label{fig_color_g}}
\end{figure*}

\begin{figure}[ht!]
	\centering 
	\includegraphics[scale=0.56]{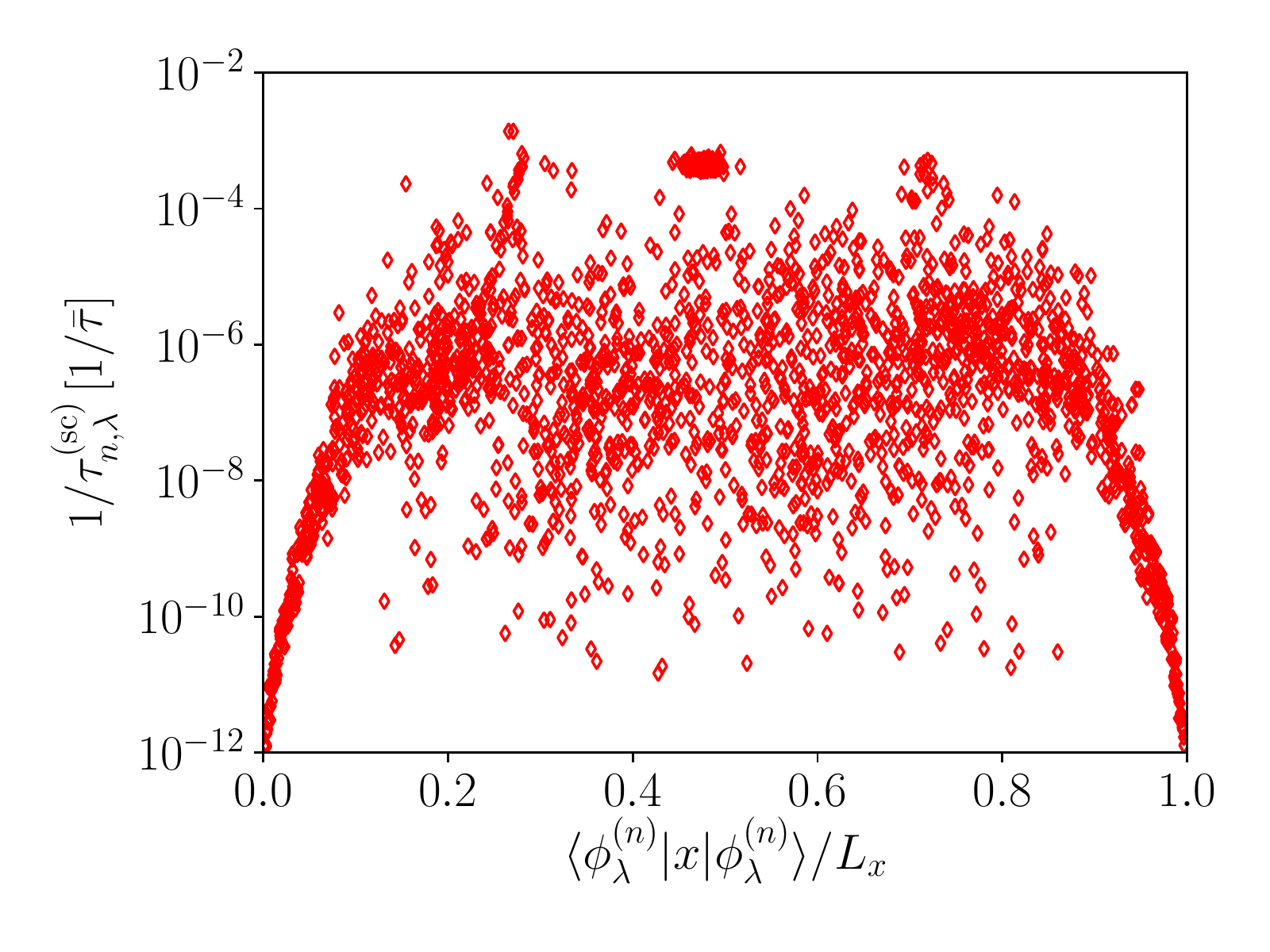}  \\
	\includegraphics[scale=0.56]{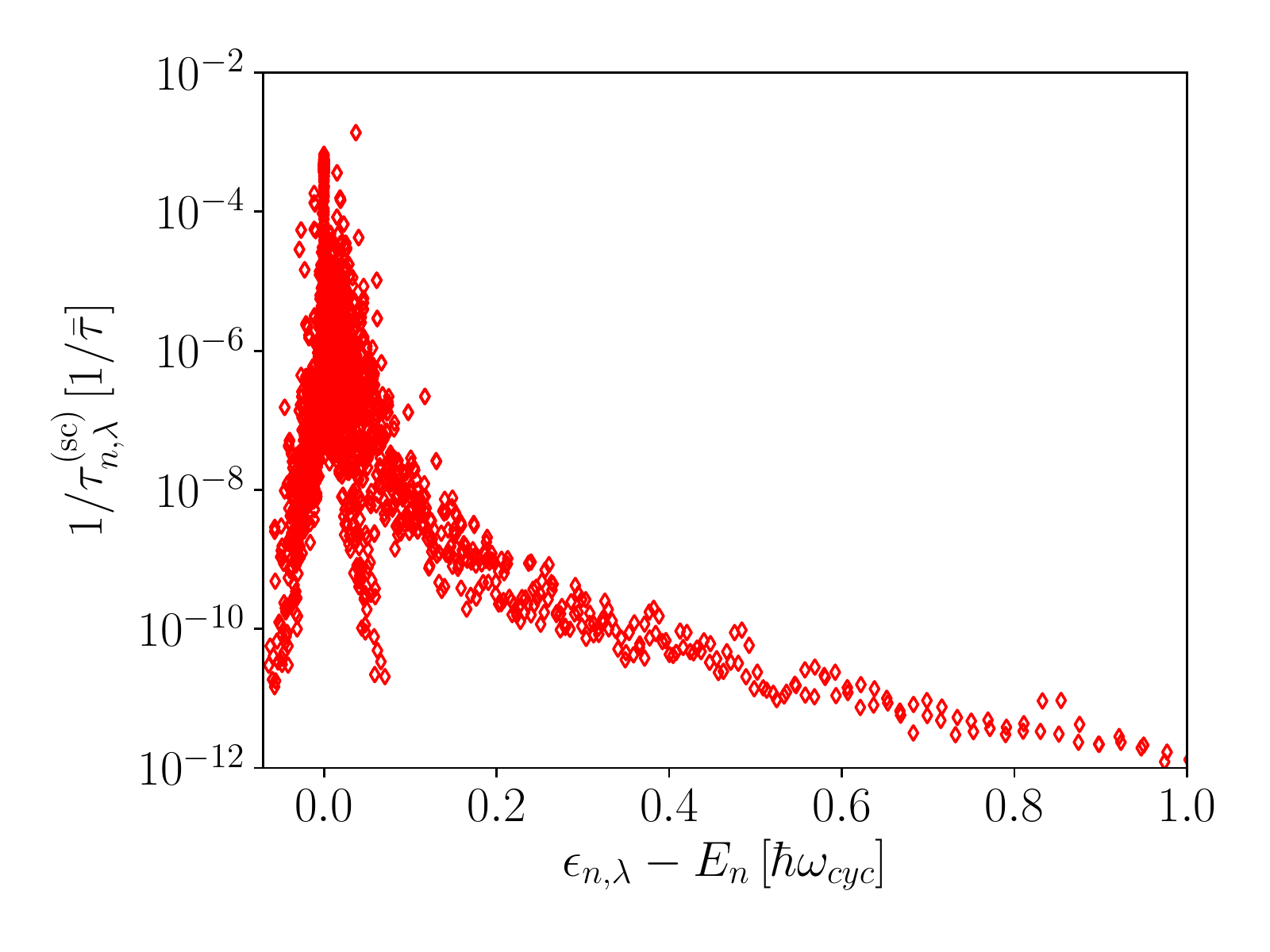} 
	\caption{Top panel: scattering rate $1/\tau^{(sc)}_{n,\lambda}$ due to cavity-mediated hopping for the disordered eigenstates $\vert \phi^{(n)}_{\lambda} \rangle$ as a function of their average position $x$, normalized by the characteristic scattering rate $1/\bar{\tau}$ defined in Eq. (\ref{tau}) and that includes the dependence on the vacuum Rabi coupling. Bottom panel: same quantity, but as a function of the energy of the disordered eigenstates. In the numerical evaluation of Eq. (\ref{rates}), the Dirac delta is represented by a Lorentzian with an energy width equal to $10^{-5} \hbar \omega_{\rm cyc}$. Other parameters are as in Fig. \ref{example_bulk}.  	\label{fig_rates}}
	
\end{figure}

\begin{figure}[ht!]
	\centering 
	\includegraphics[scale=0.57]{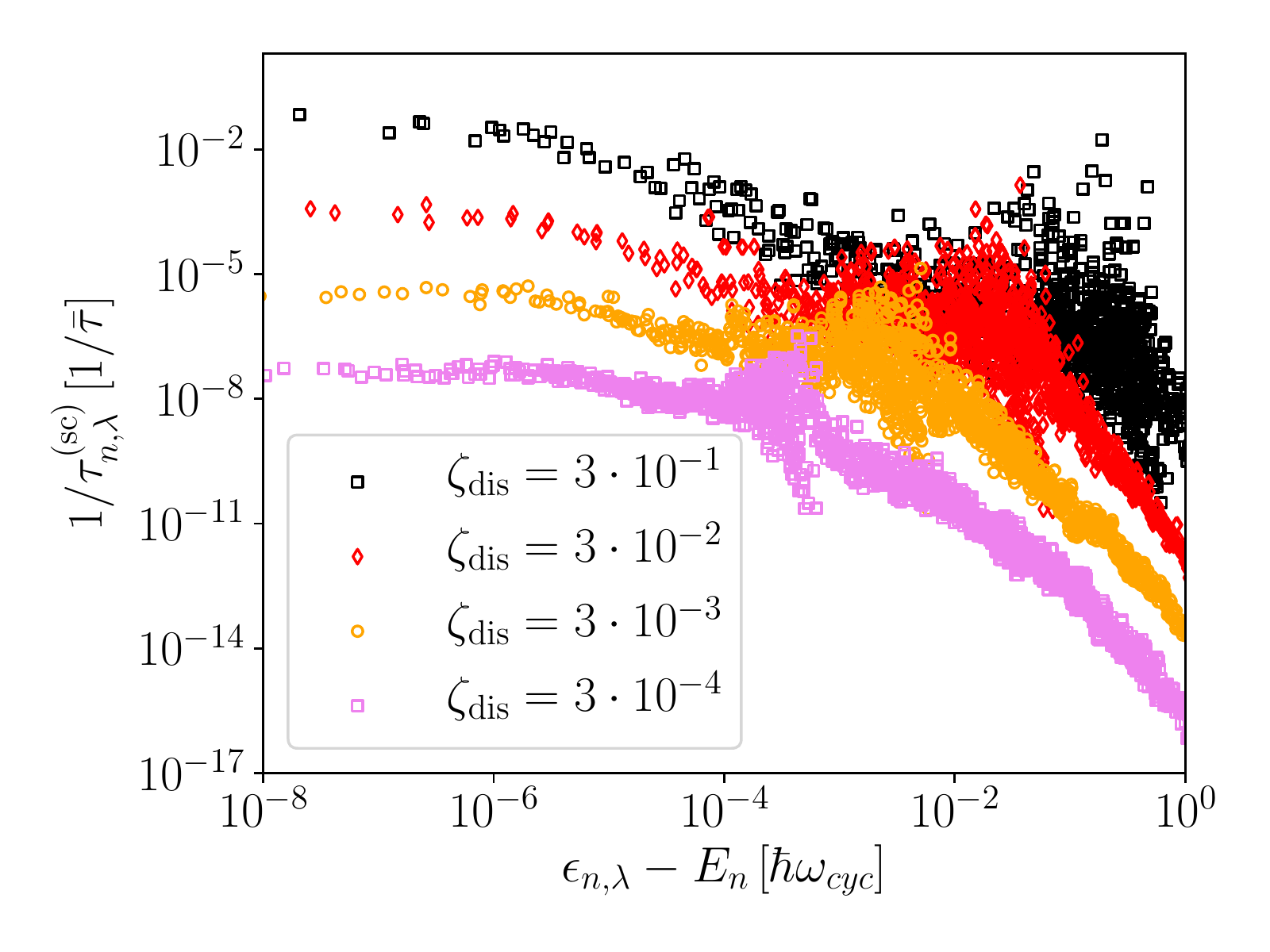}  
	\caption{Normalized scattering rate due to cavity-mediated hopping as a function of the energy difference $\epsilon_{n,\sigma} - E_n$ (log scale) for four different amplitudes of disorder, quantified by the dimensionless quantity $\zeta_{\rm dis} = N_{\rm imp}{\mathcal V}^{({\rm imp})}_{\rm max}/(\hbar \omega_{\rm cyc} L_x L_y)$: $\zeta_{\rm dis} = 3 \cdot 10^{-1}$ for the black squares, $\zeta_{\rm dis} = 3 \cdot 10^{-2}$ (as in previous figures) for red diamonds, $\zeta_{\rm dis} = 3 \cdot 10^{-3}$ for orange circles, $\zeta_{\rm dis} = 3 \cdot 10^{-4}$ for violet squares. Other parameters are as in Fig.  \ref{example_bulk}, including the number of impurities $N_{\rm imp} = 2000$ that is fixed. \label{fig_rates_disorder}}
	
\end{figure}

\begin{figure}[t!]
	\centering 
	\includegraphics[scale=0.57]{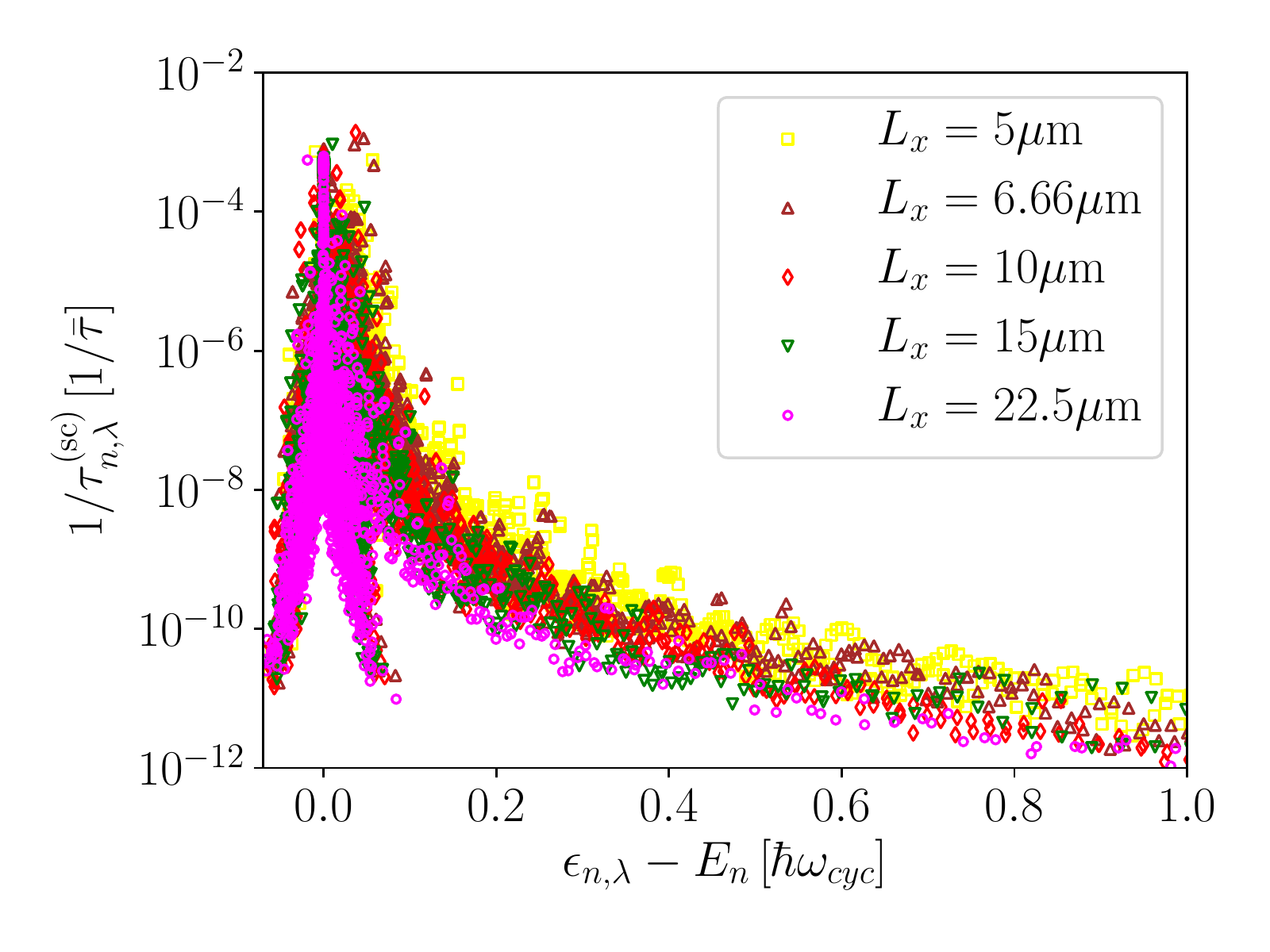}  
	\caption{Normalized scattering rate due to cavity-mediated hopping as a function of the energy difference $\epsilon_{n,\sigma} - E_n$ (log scale) for five different values of the transverse length: $L_x = 5 \mu{\rm m}$ for the yellow squares,  $L_x = 6.66 \mu{\rm m}$ for the  brown triangles, $L_x = 10  \mu{\rm m}$ for the  red diamonds (as in previous figures),   $L_x = 15 \mu{\rm m}$ for the green flipped triangles, and $L_x = 22.5  \mu{\rm m}$ for the  violet circles. 
		 Other parameters are as in Fig. \ref{example_bulk}, including the edge region width $L_{\rm e} = 2.5 \mu{\rm m}$.		\label{fig_rates_L_x}}
	
\end{figure}

\section{Finite-size numerical results}
\label{finite-size}

In this section, we report numerical results for finite-size systems, present a comprehensive study of the dependence of the cavity-mediated hopping as a function of the relevant physical quantities and determine the scaling properties.
To model single-particle electronic disorder, we have considered the sum of $N_{\rm imp} \gg 1$ randomly distributed impurity potentials, namely:
\begin{equation}
\label{impurity_pot}
V(\mathbf{r}) =  \sum_{j = 1}^{N_{\rm imp}} {\mathcal V}^{(\rm imp)}_j
\delta ({\mathbf r} -{\mathbf r}_j) \, , \end{equation}
where the $j$-th impurity random position ${\mathbf r}_j$ is  uniformly distributed in the rectangular sample of size $L_x \times L_y$. The random impurity strength ${\mathcal V}^{(\rm imp)}_j$ of the 2D Dirac delta potential $\delta ({\mathbf r} -{\mathbf r}_j)$ is uniformly distributed in the interval $[-{\mathcal V}^{(\rm imp)}_{\mathrm max},{\mathcal V}^{(\rm imp)}_{\mathrm max}]$. The corresponding matrix elements for the $n$-band are
\begin{equation}
V^{(n)}_{\kappa,\kappa'} = \sum_{j = 1}^{N_{\rm imp}} {\mathcal V}^{(\rm imp)}_j \Psi^{\star}_{n,\kappa} ({\mathbf r}_j)
\Psi_{n,\kappa'} ({\mathbf r}_j) \, .
\end{equation}

To model the wall edge potential, we have considered the function $W(x) = V_{\rm e} \tan^2 \left ( r \frac{\pi}{2} \frac{x- L_{\rm e}}{L_{\rm e}} \right ) $ for $0< x < L_{\rm e}$ (with $r$ close to 1),  $W(x) = V_{\rm e} \tan^2 \left ( r \frac{\pi}{2}\frac{x- (L_x-L_{\rm e})}{L_{\rm e}} \right ) $ for $L_x- L_{\rm e}< x < L_{x}$ and $W(x) = 0$ elsewhere in the bulk.

\label{disorder}
In Fig. \ref{example_bulk}, we report the exact diagonalization results for the single-particle disordered energy eigenvalues $\epsilon_{n,\lambda}$ as a function of $\langle \phi^{(n)}_{\lambda} \vert x \vert \phi^{(n)}_{\lambda} \rangle$, which is the expectation value of the position $x$ on the corresponding disordered eigenstates $\vert \phi^{(n)}_{\lambda} \rangle$. 
The horizontal and vertical lines indicate respectively the average $x$ position and energy of a bulk state $\lambda_{\rm b}$ taken as illustrative example.  The absolute value $\vert \tilde{g}^{(n)}_{\lambda_{\rm b},\mu} \vert$ of the vacuum Rabi frequencies  between such state $\vert \phi^{(n)}_{\lambda_{\rm b}} \rangle$ and the states 
$\vert \phi^{(n+1)}_{\mu} \rangle$ belonging to the $(n+1)$-band are reported in the middle panel of Fig. \ref{example_bulk}. Such vacuum Rabi frequencies given by Eq. (\ref{dis_g}), normalized to $\tilde{g} \sqrt{1+n}$, are plotted as a function of the average position $\langle \phi^{(n+1)}_{\mu} \vert x \vert \phi^{(n+1)}_{\mu} \rangle$ along the $x$ direction. The vacuum Rabi frequency $\tilde{g}^{(n)}_{\lambda_{\rm b},\mu}$ exhibits large random fluctuations around a mean value that is rather flat in the bulk. The coupling of the considered bulk state to $\vert  \phi^{(n+1)}_{{\mu}} \rangle$ instead collapses when the average $x$ position of $\vert  \phi^{(n+1)}_{\mu} \rangle$ approaches the edges of the sample. From the energies $\epsilon_{n,\lambda}$ and $\epsilon_{n+1,\mu}$ of the single-particle disordered eigenstates and the vacuum Rabi frequencies  $\tilde{g}^{(n,n+1)}_{\lambda,\mu}$, we can get the cavity-mediated hopping coupling energy $\vert \Gamma^{(n)}_{\lambda_{\rm b}, \lambda'}\vert$ between states $\lambda_{\rm b}$ and $\lambda'$ in the same Landau band, as shown in the bottom panel of Fig. \ref{example_bulk}.
The cavity-mediated hopping energy is normalized to the quantity
\begin{equation}
 \tilde{\Gamma}\equiv \frac{\hbar \tilde{g}^2(1+n)}{\omega_{\rm cyc} + \tilde{\omega}_{\rm cav}} \label{E_ref} \,.
 \end{equation}
 Again, we see that the bulk state $\lambda_{\rm b}$ is coupled by the cavity-mediated hopping to all the other bulk states. Note that the lone point that is much larger than the rest corresponds to the diagonal term $\vert \Gamma^{(n)}_{\lambda_{\rm b}, \lambda_{\rm b}} \vert \simeq \tilde{\Gamma}$, which is the absolute value of the second-order energy shift due to the interaction with the cavity mode. 
 
 In Fig. \ref{example_edge_1}, we plot the same quantities as in Fig. \ref{example_bulk}, but considering a given edge state labeled by the index $\lambda_{{\rm e} 1}$ and energetically close to the bulk states. We see that such an edge state is coupled by the cavity vacuum field to all the bulk states, including the edge state with the same energy on the other side of the sample, although the coupling diminishes with distance. Note that the spatial range of the interaction is much larger than the cyclotron length (for the considered parameters $l_{\rm cyc} = 0.0029 \, L_x$)
 
  In Fig. \ref{example_edge_high}, we consider an edge state labeled by the index $\lambda_{{\rm e} 2}$ and at much higher energy (no energy overlap with the bulk band). We see that the coupling is dramatically suppressed, in particular that to the opposite edge state with the same energy is suppressed by many orders of magnitude. 
  
  The dependence of $\vert \tilde{g}^{(n,n+1)}_{\lambda,\mu}\vert$ and $\vert \Gamma^{(n)}_{\lambda,\lambda'}\vert$ for every state is reported in Fig. \ref{fig_color_g}. On the left panel, a color plot (green sequential scale) of the normalized vacuum Rabi frequency is shown as a function of the average $x$ position of the state $\lambda$ in the $n$-band and the state $\mu$ in the $(n+1)$-band. On the right panel, we display a color plot (blue sequential scale) of the normalized cavity-mediated hopping as a function of the $x$ position of the $\lambda$ and $\lambda'$ states in the $n$-band. Note that the data in the middle and lower panels of Figs. \ref{example_bulk}, \ref{example_edge_1} and  \ref{example_edge_high} are cuts of the general color plots in Fig. \ref{fig_color_g} for three specific values of the state index $\lambda$. These 2D plots show in a global way the long-range nature of the cavity-mediated hopping. 

The effect of the cavity-mediated hopping can be quantified by the scattering times $\tau^{({\rm sc})}_{n,\lambda}$ obtained via the Fermi golden rule expression in Eq. (\ref{rates}). The results are reported in Fig. \ref{fig_rates} for the same parameters as the previous figures. In the top panel, the scattering rates $1/\tau^{({\rm sc})}_{n,\lambda}$ are plotted as a function of the average position  $\langle \phi^{(n)}_{\lambda} \vert x \vert \phi^{(n)}_{\lambda} \rangle$, allowing to directly distinguish between bulk and edge states. The bottom panel instead shows the same quantity as a function of the single-particle disordered energies $\epsilon_{n,\lambda}$. The scattering time $1/\tau^{({\rm sc})}_{n,\lambda}$ is expressed in units of
\begin{equation}
\label{tau}
\frac{1}{\bar{\tau}} \equiv \frac{2 \pi}{\hbar} \, \tilde{\Gamma}^2 \, \frac{N_{\rm deg}}{\hbar \omega_{\rm cyc}} \,  ,
\end{equation}
which is a characteristic rate associated to the characteristic hopping coupling $\tilde{\Gamma}$ defined earlier in Eq. (\ref{E_ref}) and to the density of states constructed in terms of the Landau degeneracy $N_{\rm deg}$ and cyclotron energy. Such a quantity can be rewritten as
\begin{equation}
\label{tau_other}
\frac{1}{\bar{\tau}}  = 2  \pi \frac{\tilde{g}^4 (1+n)^2\, N_{\rm deg}}{(\omega_{cyc}+\tilde{\omega}_{\rm cav})^2 \omega_{\rm cyc}}  
= \frac{2  \pi}{N_{\rm el}} \frac{\tilde{\Omega}^4 (1+n)^2/\nu}{(\omega_{cyc}+\tilde{\omega}_{\rm cav})^2 \omega_{\rm cyc}}
 \,.
\end{equation}

For a fixed number of electrons, the characteristic rate has a nonlinear dependence on the vacuum Rabi frequency and vanishes when the light-matter interaction tends to zero. In Fig. \ref{fig_rates} it is apparent that the cavity-mediated scattering rate is maximum in the bulk, where it fluctuates considerably. Going to the edges (see left and right side of the top panel of Fig. \ref{fig_rates}), the scattering rate decreases as well as its fluctuations. From this plot, we see that the edge states become asymptotically free (i.e., vanishing scattering rate) when they approach the sample boundaries in the $x$ direction. Equivalently, as reported in the bottom panel of Fig. \ref{fig_rates}, the scattering rate of the edge states drastically collapses when their energy increases away from the bulk band. Instead, the edge states that are energetically closer to the bulk band undergo similar scattering rates to the bulk states. 

 Fig. \ref{fig_rates_disorder} reports how the amplitude of disorder affects the cavity-induced scattering rate $1/\tau^{({\rm sc})}_{n,\lambda}$ (normalized to $1/\bar{\tau}$) versus the normalized energy difference $(\epsilon_{n,\lambda}-E_n)/\hbar \omega_{\rm cyc}$, this time represented in log scale in the interval $[10^{-6},1]$. The disorder strength is quantified by the dimensionless quantity  
 \begin{equation}
 \zeta_{\rm dis} \equiv \frac{N_{\rm imp }}{L_x L_y} \frac{{\mathcal V}^{({\rm imp})}_{\rm max}}{\hbar \omega_{\rm cyc} }\, ,
 \end{equation}
 which is simply the number of impurities per unit area weighted by the maximal strength of the delta potentials normalized to the cyclotron energy. The red diamonds correspond to the same disorder as in the previous figures, i.e., $\zeta_{\rm dis} = 3 \cdot 10^{-2}$. The orange circles correspond to a disorder amplitude $10$ times smaller ($\zeta_{\rm dis} = 3 \cdot 10^{-3}$), while the violet squares represent a disorder $100$ times smaller than for the red diamonds. The black  squares instead correspond to a disorder strength $10$ times larger ($\zeta_{\rm dis} = 3 \cdot 10^{-1}$).  Note that here the number of impurities have been fixed ($N_{\rm imp} = 2000$), so the effect is due to the increased amplitude of the delta potentials. As expected, the cavity-mediated scattering rate increases while increasing such amplitude. For energies around the center of the disordered Landau band ($\epsilon_{n,\lambda} \simeq E_n$) and for the high-energy portion of the edge states, for a given value of $\epsilon_{n,\lambda}- E_n$ the scattering increases approximately quadratically with the disorder amplitude. Instead in the intermediate region, there are larger fluctuations and scattering rates with disorder strength $\zeta_{\rm dis}$ differing by a factor of $10$ can significantly overlap, making the dependence on disorder much weaker. Note that the form and statistical properties of the disorder certainly play a role from the quantitative point of view. In particular, disordered potentials leading to the same bandwidth of the disordered Landau band can give different quantitative results for the cavity-mediated scattering rates. A comprehensive study on different classes of disorder potentials, the role of the disorder correlation length  will be interesting points to study in the future.

 \begin{figure}[t!]
 	\centering 
 	\includegraphics[scale=0.57]{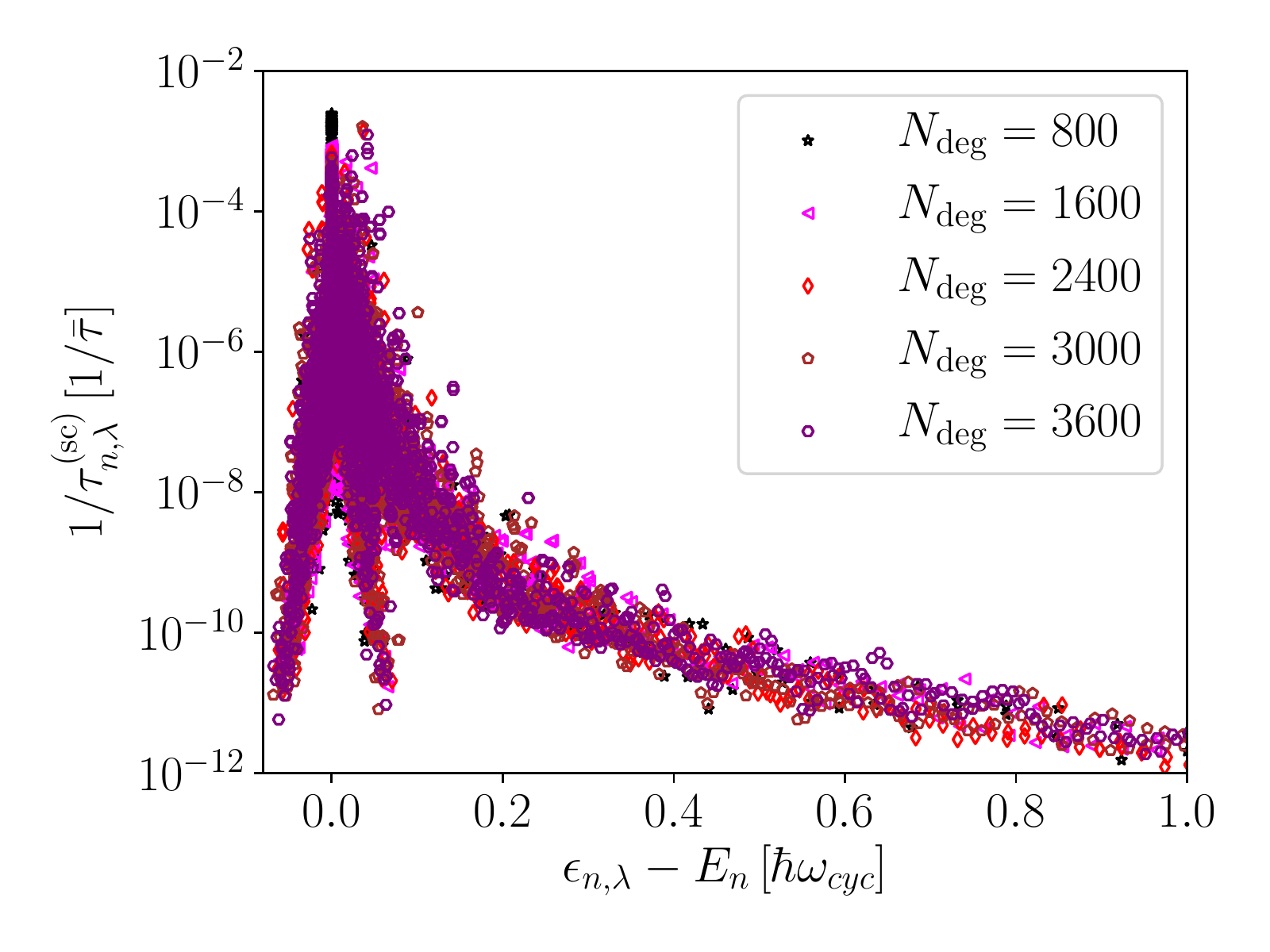}  
 	\caption{Normalized scattering rate due to cavity-mediated hopping as a function of the energy difference $\epsilon_{n,\sigma} - E_n$ for five  different values of the Landau degeneracy $N_{\rm deg}$. Note that for each value of  $N_{\rm deg}$,  the single-electron vacuum Rabi coupling $g$, the number of electrons $N_{\rm el}$ and the number of impurities $N_{\rm imp}$ are rescaled in such a way that $\Omega = g \sqrt{N_{\rm el}}$, $\nu = N_{\rm el}/N_{\rm e} $ and the dimensionless disorder strength $\zeta_{\rm dis}$ are kept constant. All other parameters are those of  Fig. \ref{example_bulk}. 	\label{fig_rates_N_deg}}
 \end{figure}

 \begin{figure}[t!]
 	\centering 
 	\includegraphics[scale=0.57]{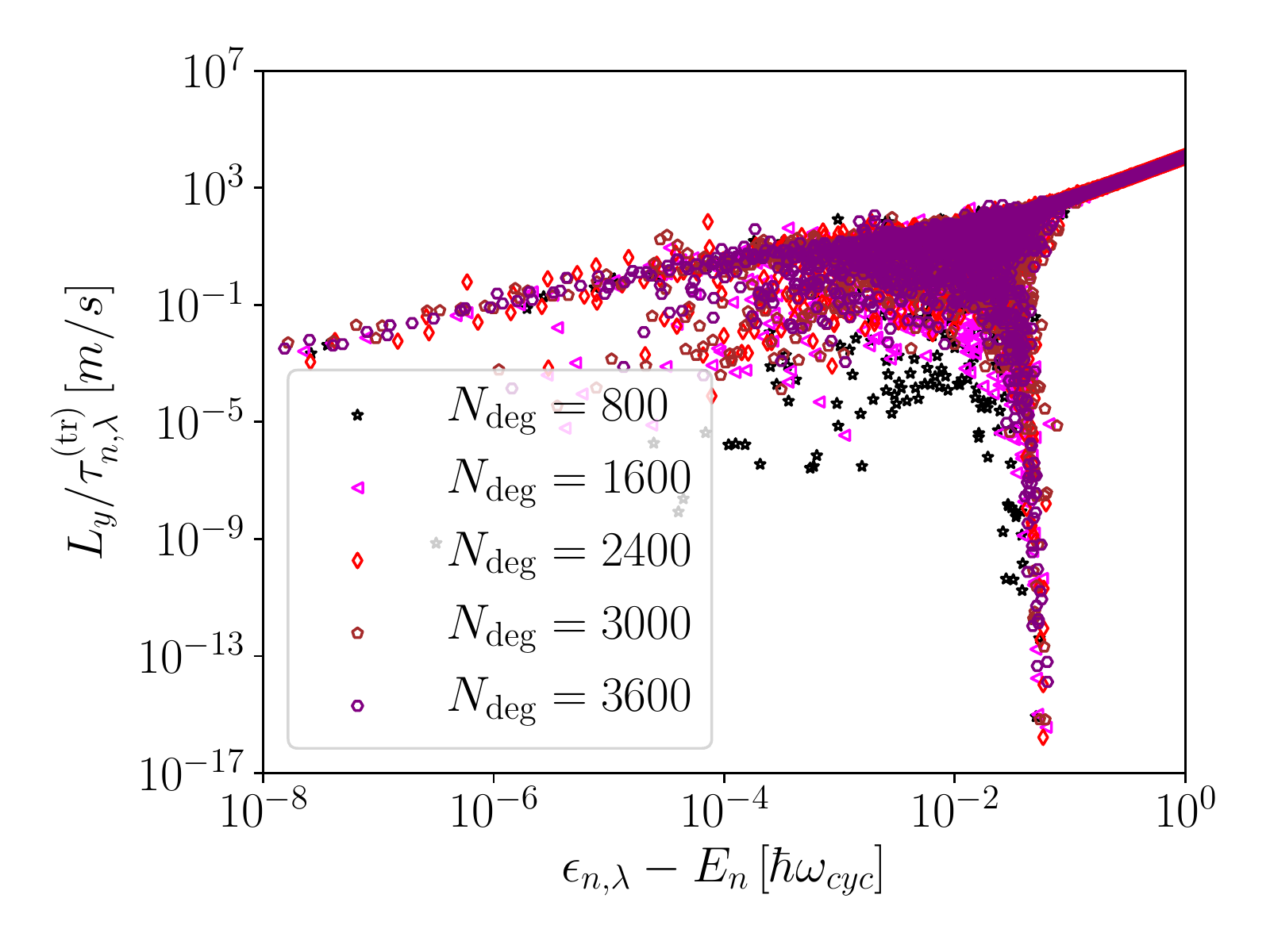}  
 	\caption{Absolute value $\vert v^{(y)}_{n,\lambda} \vert= L_y/\tau^{(tr)}_{n,\lambda}$ of the velocity  along the $y$-direction for the disordered eigenstates versus the energy difference $\epsilon_{n,\sigma} - E_n$ (log scale) for five  different values of the Landau degeneracy $N_{\rm deg}$. Same parameters as in Fig. \ref{fig_rates_N_deg}.	\label{fig_speed_N_deg}}
 \end{figure}
 
 \begin{figure}[t!]
 	\centering 
 	\includegraphics[scale=0.57]{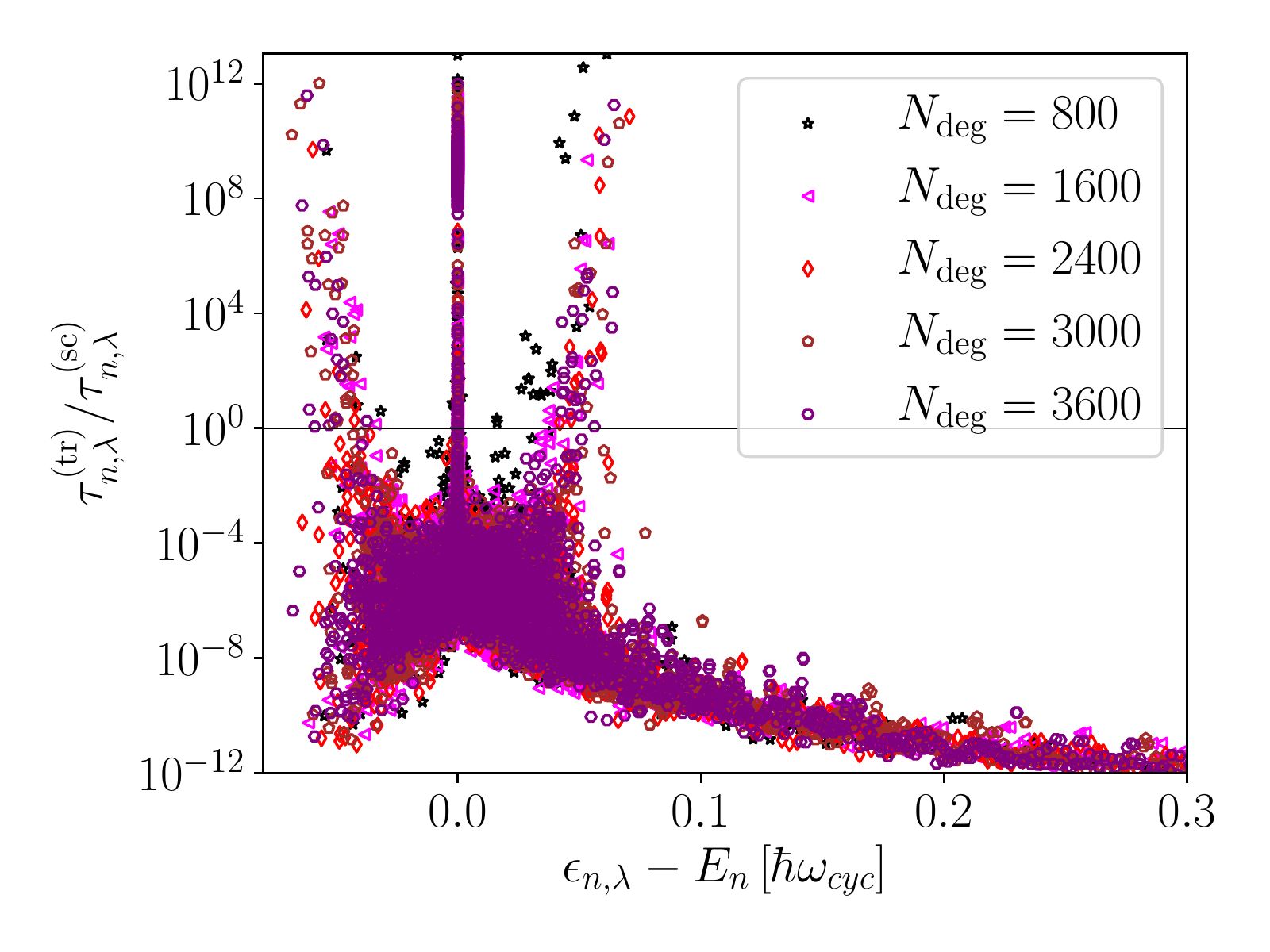}  
 	\caption{Ratio between the transit time $\tau^{(tr)}_{n,\lambda}$ and the cavity-mediated scattering time $\tau^{(sc)}_{n,\lambda}$ versus  $\epsilon_{n,\sigma} - E_n$ for five  different values of the Landau degeneracy $N_{\rm deg}$. Same parameters as in Fig. \ref{fig_rates_N_deg}.	\label{fig_transit_N_deg}}	
 \end{figure}

 The dependence on the distance between the edge walls is displayed in Fig. \ref{fig_rates_L_x}, where $L_x$ is varied between $5  \mu {\rm m}$ and $22.5 \mu {\rm m}$, while keeping all other parameters fixed, including the length $L_{\rm e} = 2.5\mu$m entering our model edge wall potential. As expected, the scattering rate for the edge states increases when $L_x$ is decreased, in particular a shortening by a factor $4.5$ translates to a reduction of approximately one order of magnitude. Of course, quantitatively this will be sensitive on the details of the wall potential and of the disorder. 
 
 It is now important to study the scaling with respect to the Landau degeneracy $N_{\rm deg}$, in particular in the 'thermodynamical' limit $N_{\rm deg} \to + \infty$.
 In Fig. \ref{fig_rates_N_deg}, \ref{fig_speed_N_deg} and \ref{fig_transit_N_deg}, we have investigated the dependence on $N_{\rm deg}$  by keeping constant all the parameters except the vacuum Rabi frequency $g$, the number of electrons $N_{\rm el}$ and the number of impurities $N_{\rm imp}$. These three quantities have been rescaled in such a way to maintain the constancy of the following quantities: $g \sqrt{N_{\rm deg}}$,  $N_{\rm el}/N_{\rm deg}$ and $\zeta_{\rm dis}$. In other words, we consider the thermodynamic limit of the system when we keep constant the collective vacuum Rabi frequency $\tilde{\Omega}$, the filling factor $\nu$ and the bandwidth of the disorder Landau band normalized to the cyclotron energy. Note also that in this limit since $L_x$ is fixed, the channel length $L_y$  increases linearly with $N_{\rm deg}$. In Fig. \ref{fig_rates_N_deg}, we report the cavity-mediated scattering rates $1/\tau^{(sc)}_{n,\lambda}$ normalized to the characteristic scattering rate $1/\bar{\tau}$ defined in Eq. (\ref{tau}) and (\ref{tau_other}).
Fig. \ref{fig_rates_N_deg} shows that such a quantity versus energy converges with increasing value of $N_{\rm deg}$, confirming the fact that $1/\bar{\tau}$, which crucially contains  the dependence on the vacuum Rabi frequency, is the relevant scattering rate.

The absolute value $\vert v^{(y)}_{n,\lambda} \vert$ of the velocity of the disordered eigenstates  is reported in Fig. \ref{fig_speed_N_deg} as a function of energy, showing an analogous convergence by increasing enough $N_{\rm deg}$. Such a quantity is the ratio between the channel length $L_y$ and the transit time $\tau^{(tr)}_{n,\lambda}$. Since in the considered limit $L_y \propto N_{\rm deg}$, then also $\tau^{(\rm tr)}_{n, \lambda} \propto N_{\rm deg}$. The high-speed edge states are easily  recognizable at high energies. Their speed decreases by many orders of magnitude when their energy difference with respect to $E_n$ (the central energy of the Landau band) tends to 0. A second branch of slow states corresponds to the localized states in the lower and higher energy tails of the disordered bulk Landau band.

Finally, Fig. \ref{fig_transit_N_deg} reports crucial results with the dependence of the ratio between the transit time $\tau^{(\rm tr)}_{n,\lambda}$ and the cavity-mediated scattering time  $\tau^{(\rm sc)}_{n,\lambda}$ versus energy. Again, we see clearly that for increasing $N_{\rm deg}$ the points superimpose, showing that the thermodynamic limit is already well approached when $N_{\rm deg}$ is of the order of one thousand. Importantly, when the scattering time becomes comparable or larger to the transit time, the transport properties of the Landau band are expected to be affected. We see that the effect is most significant for the slow edge states and for the localized bulk states in the energy tails of the Landau band.    

\section{Discussion and relation to quantum Hall transport}
\label{Discussion} 
The scattering created by cavity-mediated hopping can be relevant for quantum Hall transport \cite{Appugliese_2021} due to the long-range nature of the effect (i.e., occurring on a scale much larger than the cyclotron length $l_{\rm cyc}$).  Indeed, the cavity introduces an additional scattering mechanism for the bulk states, but most importantly it creates a coupling between the edge states and the bulk. For edge states energetically close to the bulk, we have just seen above that there can be also a direct coupling to the opposite edge state, as observed in \cite{Appugliese_2021}. Note that this inter-edge coupling can also be enhanced via incoherent multiple scattering processes occurring in the bulk. 

As a consequence, the cavity-mediated long-range scattering can be a source of deviation from the metrological quantization of the integer quantum Hall plateaus. In the edge picture of the integer quantum Hall effect \cite{Buttiker1988,Girvin_2019}, the bulk states are insulating, while the current is carried by the chiral edge states corresponding to the classical skipping orbits (see sketch in Fig. \ref{sketch}). The quantization is due to the absence of back-scattering for the chiral edge channels. The cavity-mediated hopping can create an effective coupling between opposite edges and threaten the topological protection of the quantum Hall effect.

Note that, even if the cavity-mediated interaction conserves the spin, there must be a different impact on the quantum Hall effect for different spin channels. Indeed, the odd integer filling factor plateaus (around $\nu = 2n +1$) are associated to the $n$-band with spin $\sigma = \, \uparrow$, while the even integer plateaus (around $\nu = 2n+2$) are associated to the $n$-band with spin $\sigma = \, \downarrow$ (see Fig. \ref{sketch}). For an odd integer plateau, the corresponding edge states responsible for the quantized Hall conductance have an energy $E^{({\rm odd})}_{\rm edge}$ such that $E^{({\rm odd})}_{\rm edge}-E_{n,\uparrow} \in [0, {\rm g}_{\rm e} \mu_{\rm B} B]$. Instead, for an even integer plateau, the energy of the corresponding edge states is such that $E^{({\rm even})}_{\rm edge}-E_{n,\downarrow} \in [0, \hbar \omega_{\rm cyc}-{\rm g}_{\rm e} \mu_{\rm B} B]$.
Since the edge states become less affected by the cavity-mediated hopping when the energy difference from the corresponding bulk states increases (as shown in Fig. \ref{fig_rates}), the even integer plateaus must be less affected than the odd ones when $\hbar \omega_{\rm cyc} \gg {\rm g}_{\rm e} \mu_{\rm B} B$, which is the case in GaAs 2D electron gases \cite{Appugliese_2021}.

\section{Conclusions and perspectives}
\label{conclusions}

In this article, we have presented a detailed microscopic theory showing how in a disordered quantum Hall system the coupling to the vacuum fields of a cavity resonator can mediate an effective long-range hopping between  single-particle disordered eigenstates. The mechanism is due to the counter-rotating (anti-resonant) terms of the quantum light-matter interaction. For an electron in the last occupied Landau band with orbital quantum number $n$ and spin $\sigma$, it involves a macroscopic number of intermediate states consisting of one virtual cavity photon and one electron in the  $(n+1)$-band with the same spin $\sigma$. Due to its non-local and vacuum nature (no illumination), this effect can intrinsically weaken the topological protection of the integer quantum Hall states, because it can create a coupling between opposite edge states, which are normally decoupled. By accounting both for paramagnetic coupling and diamagnetic renormalization, we have determined the effective hopping in terms of the single-particle eigenstates in presence of a random disorder potential and a wall potential at the edges. Moreover, we have studied the corresponding scattering rates for the disordered eigenstates, obtained at the level of Fermi golden rule. We have also individuated the scaling properties and the relevant intensive quantities. The theory shows that the cavity-mediated hopping affects both bulk and edge states. The effect for edge states increases when their energy approaches that of the bulk band. Instead, at high energy the edge states become asymptotically free. 

From the theoretical point of view, future possible developments  encompass the generalization to spatially non-homogeneous photon modes and multimode cavities. Since the mechanism is anti-resonant, the presence of other modes can enhance the cavity-mediated hopping, because it introduces additional intermediate states. Note that the spatial inhomogeneity of a mode, especially near the edges, can also produce a similar effect to the electronic disorder. An other interesting development is a microscopic description of transport by considering the cavity-mediated scattering dynamics on the disordered eigenstates including Pauli blocking, multiple scattering processes and applied electrical biases in multiprobe geometries, as well as the competition with other scattering mechanisms. Moreover, the role of cavity quantum fields on the quantum Hall effects with 2D van der Waals materials \cite{Ajayan2016} involving the interplay with additional degrees of freedom (valley, layer) or on other flat band systems are certainly intriguing directions to investigate. With the recent experimental observation of electron scattering by vacuum fields in quantum Hall systems \cite{Appugliese_2021}, the experimental and theoretical research on the control of materials by cavity vacuum fields \cite{Garcia_2021} is destined to considerably accelerate and expand.

\acknowledgements{I would  like to warmly thank Felice Appugliese, Josefine Enkner, Giacomo Scalari and J\'er\^ome Faist for the numerous stimulating discussions and for showing their experimental results \cite{Appugliese_2021} prior to publication. I wish also to thank Zakari Denis for helpful tips about the Julia programming language.
I would additionally like to acknowledge support from the ANR project TRIANGLE (ANR-20-CE47-0011) and from the FET FLAGSHIP Project PhoQuS (grant agreement ID no. 820392)}.
 \bibliography{paper_cavity_QH.bib}
\clearpage
\end{document}